\documentclass[showkeys,amsmath,amssymb,nofootinbib,superscriptaddress]{revtex4}

\usepackage{graphicx}
\usepackage{epsfig}
\usepackage{amsfonts}
\usepackage{amssymb}
\usepackage{color}
\usepackage{float}
\usepackage{amsmath}                  
\usepackage{hyperref}
\usepackage{lscape}
\usepackage{ulem}
\usepackage{xcolor}
\usepackage{tensor}
\usepackage{soul}						
\usepackage{booktabs} 						
\usepackage{multirow}						
\usepackage{soul} 						
\usepackage{microtype}		
\usepackage{mathpazo}
\usepackage{newpxtext}

\renewcommand{\em}{\it}

\begin{document}

\title{Dark Gravitational Field on Riemannian and Sasaki~Spacetime}

\begin{abstract}
	The aim of this paper is to provide the geometrical structure of a gravitational field that includes the addition of dark matter in the framework of a Riemannian and a Riemann--Sasaki spacetime. By means of the classical Riemannian geometric methods we arrive at modified geodesic equations, tidal forces, and Einstein and Raychaudhuri equations to account for extra dark gravity. We further examine an application of this approach in cosmology. Moreover, a possible extension of this model on the tangent bundle is studied in order to examine the behavior of dark matter in a unified geometric model of gravity with more degrees of freedom. Particular emphasis shall be laid on the problem of the geodesic motion under the influence of dark matter.
\end{abstract}
% Keywords

\author{Panayiotis Stavrinos}
\email{pstavrin@math.uoa.gr}
\affiliation{Department of Mathematics, National and Kapodistrian University of Athens, Athens, 15784, Greece}

\author{Christos Savvopoulos}
\email{sph1400190@uoa.gr}
\affiliation{Department of Physics, National and Kapodistrian University of Athens, Athens, 15784, Greece}
%%%%%%%%%%%%%%%%%%%%%%%%%%%%%%%%%%%%%%%%%%
\maketitle

\section{Introduction}\label{section1}
Recent advances in the theoretical and observational field of cosmology have shown the significance of the existence of dark matter and dark energy~\cite{Farnes, Fermilab, mediator, data, Royal, Riess1, arxiv, Symmetry, Valentino1, Leib, Leib2, Krauss, Saul1, Riess2, Abbott, Yijung, Sarkar, Valentino2, Leib3, Saul2, Velten, Barger, Diamanti, Valentino3, GR}. There is considerable evidence that most of the mass in the universe is neither in the luminous matter in galaxies, nor in the radiation detected so far. Mass can be detected by its gravitational influence even if it cannot be seen directly~\cite{Hartle}. Evidence for the existence of dark matter comes for example from the study of gravitational lensing, cosmic microwave background radiation (CMB) or the rotational curves of spiral galaxies~\cite{Godlowski,Salucci}.

According to observational results, dark matter plays a dominant role in the field of evolution and acceleration of the universe. There are many pieces of evidence that the visible matter and detectable radiation comprise only a small fraction of the mass in the universe, perhaps as little as a few percent~\cite{Hartle}. Therefore the study of dark matter is central for cosmology. A~geometric model for gravity with dark matter based on a spacetime manifold endowed with a Riemannian metric is necessary, because~of the existence of contributions to gravity, curvature, tidal forces etc. due to dark matter whose gravitational interaction could be described using a Riemmanian geometric framework similar to that of ordinary matter as shall be assumed throughout this~work. 

These contributions can be observed in the velocities of nearby trajectories (geodesics) of particles (or clusters thereof, e.g.,~planets), in~the orbital motions of member galaxies in galaxy clusters or in the relative change of velocity of the deviation vector which shall now be altered to account for the existence of extra ``dark'' gravity. The~above mentioned combined with observational data give rise to a question regarding whether or not one accepts the assumption that test bodies move on geodesics of a given inertial metric of the base spacetime~\cite{Harko1,Harko2}; in other words, whether or not the geodesics remain the same during the evolution of the universe with the addition of dark~matter. 

In a more general geometric framework, the~tangent bundle of a spacetime manifold, can extend the limits of a ``unified'' gravitational field in more than four dimensions. The~development of the metric geometry of tangent bundles first began with the introduction of the natural Sasaki metric, in~the fundamental paper of Sasaki published in 1958~\cite{Sas-58}, which we shall assume for a further geometric model with dark matter. This choice proves crucial as the tangent bundle allows for a gravitational field with more degrees of freedom. Dark matter could cause extra gravitational influence on all scales, which can be illustrated well in extra dimensions~\cite{Mukhanov, Nature, Question, Brazil, Group}. A~geometric frame that potentially materializes this concept is, therefore, that of Sasaki, where the underlying metric structure is~Riemannian.

This work is organized as follows: in Section~\ref{section2}, we examine the gravitational field with the addition of extra dark gravity in a Riemannian spacetime setting. In~particular, in~Section~\ref{parA} we provide the geodesic equations and their deviation, in~Section~\ref{parB} we study the Einstein equations and in Section~\ref{parC} the Raychaudhuri equation. In~addition, in~Section~\ref{parapp} we derive the Friedmann equations and the continuity equation as an application using the $F(R)$ gravity model for dark matter. In~Section~\ref{section3}, we extend our study of the dark and ordinary gravitational field on the tangent bundle of a Sasaki spacetime. In~Section~\ref{parD} we derive the equations of geodesic deviation on the Sasaki tangent bundle, and~in Section~\ref{parE} we provide the foundations for an extension of dark gravity on the tangent bundle. Finally, in~Section~\ref{conclusion} we summarize the results of this work and in Appendix \ref{App} we present some further geometric~results.

\section{Dark Gravity in the Riemannian~Spacetime}\label{section2}
Let us consider a (pseudo-)Riemannian 4-dimensional spacetime $M$ containing both matter and dark matter equipped with a metric $g$. For~the purposes of our study, an~additive relation for the metric tensor is assumed, such that the contributions to the metric of ordinary and dark matter can be viewed~separately. 

Let
\begin{equation}\label{lineel}
ds^2=\tensor{g}{_{ij}}(x)dx^idx^j
\end{equation}
be the metric of this 4-dimensional spacetime, where we assume that the unified metric
\begin{equation}\label{assumption}
\tensor{g}{_{ij}}(x)=\tensor{g}{^{(O)}_{ij}}(x)+\tensor{g}{^{(D)}_{ij}}(x)
\end{equation}
where $\tensor{g}{^{(O)}_{ij}}(x)$ is the sectoral metric of ordinary matter and $\tensor{g}{^{(D)}_{ij}}(x)$ that of dark matter\footnote{For all three metric tensors a metric signature ($-$,+,+,+) shall be assumed in their respective space.}.

Due to the geometry of the space that has been chosen, we have to use the unified metric tensor, $\tensor{g}{_{ij}}$, for~such operations as raising, lowering and contracting indices of tensors. Since, in~our study, we deal with tensors related to ordinary or dark matter's spacetime, we shall attempt to relate such concepts within the framework of the unified geometric space. From~a physical perspective, a~unified framework of gravity which includes the gravitational interaction of ordinary and dark matter is necessary in order to describe the gravitational effects of the large scale universe structures (e.g.,~to~explain the rotation of galaxies or the motions of clusters). 

\subsection{Geodesics and Tidal~Forces}\label{parA}
First, we find that the Chistoffel symbols of first kind are:
\begin{equation}
\tensor{\Gamma}{_i_j_k}=\tensor{\Gamma}{^{(O)}_i_j_k}+\tensor{\Gamma}{^{(D)}_i_j_k}
\end{equation}
where,
\begin{equation}
\tensor{\Gamma}{^{(O)}_i_j_k}=\frac{1}{2}\bigg(\frac{\partial \tensor{g}{^{(O)}_i_j}}{\partial x^k}+\frac{\partial \tensor{g}{^{(O)}_i_k}}{\partial x^j}-\frac{\partial \tensor{g}{^{(O)}_j_k}}{\partial x^i}\bigg)
\end{equation}
\begin{equation}
\tensor{\Gamma}{^{(D)}_i_j_k}=\frac{1}{2}\bigg(\frac{\partial \tensor{g}{^{(D)}_i_j}}{\partial x^k}+ \frac{\partial \tensor{g}{^{(D)}_i_k}}{\partial x^j}-\frac{\partial \tensor{g}{^{(D)}_j_k}}{\partial x^i}\bigg)
\end{equation}
are the Christoffel symbols of first kind of a space occupied exclusively by ordinary or dark matter, respectively\footnote{One must be careful that $\tensor{\Gamma}{^{(O)}_i_j_k}$ and $\tensor{\Gamma}{^{(D)}_i_j_k}$ do not function as Christoffel symbols for the unified space $(M,g)$.}.

The geodesics will then be given by the well-known equation, which needs to be modified to account for the existence of dark matter as
\begin{equation}\label{rgeod}
\frac{d^2x^i}{dt^2}+(\tensor{\Gamma}{^{(O)}^i_j_k}+\tensor{\Gamma}{^{(D)}^i_j_k}+\tensor{\gamma}{^i_j_k})\frac{dx^j}{dt}\frac{dx^k}{dt}=0
\end{equation}
since the Christoffel symbols of second kind shall have the following form:
\begin{equation}\label{Christoff}
\tensor{\Gamma}{^i_j_k}=\tensor{\Gamma}{^{(O)}^i_j_k}+\tensor{\Gamma}{^{(D)}^i_j_k}+\tensor{\gamma}{^i_j_k}
\end{equation}
where $t$ is an affine parameter and
\begin{equation}
\tensor{\Gamma}{^{(O)}^i_j_k}=\frac{1}{2}\tensor{g}{^{(O)}^i^l}\bigg(\frac{\partial \tensor{g}{^{(O)}_l_j}}{\partial x^k}+\frac{\partial \tensor{g}{^{(O)}_l_k}}{\partial x^j}-\frac{\partial \tensor{g}{^{(O)}_j_k}}{\partial x^l}\bigg)
\end{equation}
\begin{equation}
\tensor{\Gamma}{^{(D)}^i_j_k}=\frac{1}{2}\tensor{g}{^{(D)}^i^l}\bigg(\frac{\partial \tensor{g}{^{(D)}_l_j}}{\partial x^k}+ \frac{\partial \tensor{g}{^{(D)}_l_k}}{\partial x^j}-\frac{\partial \tensor{g}{^{(D)}_j_k}}{\partial x^l}\bigg)
\end{equation}
are the Christoffel symbols of ordinary and dark matter's metric contribution, respectively, and~\begin{equation}
\tensor{\gamma}{^i_j_k}=\tensor{\Gamma}{^i_j_k}-(\tensor{\Gamma}{^{(O)}^i_j_k}+\tensor{\Gamma}{^{(D)}^i_j_k})
\end{equation}
is the interaction part of the Christoffel symbols since it could be shown that $\tensor{\gamma}{^i_j_k}$ represents the interaction between the ordinary and dark matter gravitational potentials and their respective intensities\footnote{These $\tensor{\gamma}{^i_j_k}$ symbols could be explicitly calculated using~\cite{Miller}.}. 

In view of relation %We confirm thagt this is the correct relation we have used for the following.
(\ref{rgeod}), one can see that for a test particle $m_i$ moving along geodesics in the spacetime of our model the following relation holds:
\begin{equation}\label{testp}
m_i\bigg(\frac{d^2x^i}{dt^2}+\tensor{\Gamma}{^{(O)}^i_j_k}\frac{dx^j}{dt}\frac{dx^k}{dt}\bigg)=-m_i\bigg(\tensor{\Gamma}{^{(D)}^i_j_k}+\tensor{\gamma}{^i_j_k}\bigg)\frac{dx^j}{dt}\frac{dx^k}{dt}=F^i_{(D)}
\end{equation}

Seemingly, the~left-hand side of Equation~(\ref{testp}) does not give a geodesic in the Riemannian sense since the right-hand side does not vanish unless we consider dark matter to be absent. The~dark gravitational field and its interaction with ordinary matter influences the geodesics in a dominant way and give rise to a dark pseudo-force field $F^i_{(D)}$ which can be interpreted as a gravitational source for inertial force fields of interaction by an observer who does not take into account the existence of dark matter in his considerations. The~addition of dark matter also influences the curvature tensor of the unified space, $\tensor{R}{^a_b_c_d}$, which assumes the form shown in the following relation:
\begin{equation}\label{riemanncurvtensor}
\tensor{R}{^a_b_c_d}=\tensor{R}{^{(O)}^a_b_c_d}+\tensor{R}{^{(D)}^a_b_c_d}+ \tensor{r}{^a_b_c_d}
\end{equation}
where,
\begin{equation}
\tensor{R}{^{(O)}^a_b_c_d}=\frac{\partial \tensor{\Gamma}{^{(O)}^a_b_d}}{\partial x^c}-\frac{\partial \tensor{\Gamma}{^{(O)}^a_b_c}}{\partial x^d}+\tensor{\Gamma}{^{(O)}^a_c_e}\tensor{\Gamma}{^{(O)}^e_b_d}-\tensor{\Gamma}{^{(O)}^a_d_e}\tensor{\Gamma}{^{(O)}^e_b_c}
\end{equation}
is ordinary matter's Riemann tensor,
\begin{equation}
\tensor{R}{^{(D)}^a_b_c_d}=\frac{\partial \tensor{\Gamma}{^{(D)}^a_b_d}}{\partial x^c}-\frac{\partial \tensor{\Gamma}{^{(D)}^a_b_c}}{\partial x^d}+\tensor{\Gamma}{^{(D)}^a_c_e}\tensor{\Gamma}{^{(D)}^e_b_d}-\tensor{\Gamma}{^{(D)}^a_d_e}\tensor{\Gamma}{^{(D)}^e_b_c}
\end{equation}
is dark matter's curvature tensor and,
\begin{equation}
\begin{split}
\tensor{r}{^a_b_c_d}=\frac{\partial \tensor{\gamma}{^a_b_d}}{\partial x^c}-\frac{\partial \tensor{\gamma}{^a_b_c}}{\partial x^d}+\tensor{\Gamma}{^{(O)}^a_e_{[c}}\tensor{\Gamma}{^{(D)}^e_{d]}_b}+\tensor{\Gamma}{^{(D)}^a_e_{[c}}\tensor{\Gamma}{^{(O)}^e_{d]}_b}+\tensor{\Gamma}{^{(O)}^a_e_{[c}}\tensor{\gamma}{^e_{d]}_b} \\+\tensor{\Gamma}{^{(D)}^a_e_{[c}}\tensor{\gamma}{^e_{d]}_b}+\tensor{\gamma}{^a_e_{[c}}\tensor{\Gamma}{^{(O)}^e_{d]}_b}+\tensor{\gamma}{^a_e_{[c}}\tensor{\Gamma}{^{(D)}^e_{d]}_b}+\tensor{\gamma}{^a_e_{[c}}\tensor{\gamma}{^e_{d]}_b}    
\end{split}
\end{equation}
is the interaction part of the Riemann tensor\footnote{The lower-indices $[cd]$ denote an antisymmetrization (similarly, indices between parentheses shall denote symmetrization).}$^,$\footnote{One must be careful that for the unified space $(M,g)$ there is but one curvature tensor; the unified $\tensor{R}{^a_b_c_d}$.}. 

Similarly, the~equation of geodesic deviation needs to be modified, as~well. It will then take the following form:
\begin{equation}\label{rdev}
\delta_t(\delta_t W^a)=-(\tensor{R}{^{(O)}^a_b_c_d}+\tensor{R}{^{(D)}^a_b_c_d}+\tensor{r}{^a_b_c_d})V^bW^cV^d
\end{equation}
where,
\begin{equation}
\delta_t W^a=\frac{\partial W^a}{\partial t}+(\tensor{\Gamma}{^{(O)}^a_b_c}+\tensor{\Gamma}{^{(D)}^a_b_c}+\tensor{\gamma}{^a_b_c})W^bV^c
\end{equation}
$V^i=\partial_t x^i$, $W^i=\partial_s x^i$ and $x^i_s(t)$ is a two-parameter geodesic family with $t$ being the affine parameter and $s$ the selector parameter~\cite{Hartle}.

In view of relations (\ref{riemanncurvtensor}) and (\ref{rdev}), the~extra terms of the product of the Christoffel symbols of ordinary and dark matter represent the coupling of the intensities of their respective gravitational fields, i.e.,~an interaction which has been instigated by $\tensor{\gamma}{^i _{jk}}$ and propagated by the curvature tensor in the deviation of geodesics. This shows that the interaction between ordinary and dark matter has been incorporated in the geometry of the spacetime due to curvature being an intrisic geometric property of space. Thus, the~interaction of the gravitational field intensities is now being manifested in the Deviation Equation ~(\ref{rdev}) and therefore in the tidal forces. This means that the geodesics we understand are deviated by the unified curvature including both dark and ordinary matter, instead of the geodesic motion in the ordinary spacetime as we have already noticed in rel. (\ref{testp}).

\subsection{Einstein~Equations}\label{parB}
We shall now present the Riemann curvature tensor in its covariant form since it shall prove useful later on. By~lowering the upper index in Equation~(\ref{riemanncurvtensor})\footnote{\label{imp}For such operations we must always use the unified metric.} we get:
\begin{equation}\label{riemanncurvtensorcov}
\tensor{R}{_{ab}_c_d}=\tensor{R}{^{(O)}_{ab}_c_d}+\tensor{R}{^{(D)}_{ab}_c_d}+ \tensor{\varrho}{_{ab}_c_d}
\end{equation}
where,
\begin{equation}
\tensor{R}{^{(O)}_{ab}_c_d}=\tensor{g}{^{(O)}_a_e}\tensor{R}{^{(O)}^e_b_c_d}
\end{equation}
is ordinary matter's covariant Riemann tensor,
\begin{equation}
\tensor{R}{^{(D)}_{ab}_c_d}=\tensor{g}{^{(D)}_a_e}\tensor{R}{^{(D)}^e_b_c_d}
\end{equation}
is dark matter's covariant curvature tensor and,
\begin{equation}
\tensor{\varrho}{_{ab}_c_d}=\tensor{g}{^{(D)}_a_e}\tensor{R}{^{(O)}^e_b_c_d}+\tensor{g}{^{(O)}_a_e}\tensor{R}{^{(D)}^e_b_c_d}+\tensor{g}{_a_e}\tensor{r}{^e_b_c_d}
\end{equation}
is the interaction part of the~tensor.

By virtue of Equation~(\ref{riemanncurvtensorcov}), we can now find the Ricci curvature:
\begin{equation}\label{riccicurvtensor}
\tensor{R}{_{ab}}=\tensor{R}{^{(O)}_{ab}}+\tensor{R}{^{(D)}_{ab}}+ \tensor{\xi}{_{ab}}
\end{equation}
where,
\begin{equation}
\tensor{R}{^{(O)}_{ab}}=\tensor{g}{^{(O)}^c^d} \tensor{R}{^{(O)}_c_a_d_b}
\end{equation}
is ordinary matter's Ricci curvature tensor,
\begin{equation}
\tensor{R}{^{(D)}_{ab}}=\tensor{g}{^{(D)}^c^d} \tensor{R}{^{(D)}_c_a_d_b}
\end{equation}
is dark matter's Ricci curvature tensor and,
\begin{equation}
\tensor{\xi}{_{ab}}=\tensor{R}{_{ab}} - (\tensor{R}{^{(O)}_{ab}}+\tensor{R}{^{(D)}_{ab}})
\end{equation}
is the interaction part of the Ricci~tensor.

Using the form of the Ricci tensor given in Equation~(\ref{riccicurvtensor}), we can find the Ricci scalar:
\begin{equation}\label{riccicurvscalar}
R=R^{(O)}+R^{(D)}+ \kappa
\end{equation}
where,
\begin{equation}
R^{(O)}=\tensor{g}{^{(O)ab}} \tensor{R}{^{(O)}_{ab}}
\end{equation}
is ordinary matter's Ricci scalar,
\begin{equation}
R^{(D)}=\tensor{g}{^{(D)ab}} \tensor{R}{^{(D)}_{ab}}
\end{equation}
is dark matter's Ricci scalar and,
\begin{equation}
\kappa=R-(R^{(O)}+R^{(D)})
\end{equation}
is the interaction part of the Ricci~scalar.

The actions of the sectoral ordinary and dark matter gravitational fields and matter are, respectively:
\begin{equation}
S_O=\int   \bigg( \frac{1}{16\pi} R^{(O)} + \mathcal{L}_{M_O}\bigg)  \sqrt{-g^{(O)}} \, d^4x
\end{equation}
and
\begin{equation}
S_D=\int   \bigg( \frac{1}{16\pi} R^{(D)} + \mathcal{L}_{M_D}\bigg)  \sqrt{-g^{(D)}} \, d^4x
\end{equation}
where $\mathcal{L}_{M_O}$ and $\mathcal{L}_{M_D}$ describe the sectoral ordinary and dark matter fields, $g^{(O)}=\det\tensor{g}{^{(O)}_{ij}}$ and $g^{(D)}=\det\tensor{g}{^{(D)}_{ij}}$, $M_O$ and $M_D$ are the ordinary and dark matter masses, respectively\footnote{We will refrain from using specific Lagrangians neither for the ordinary nor for the dark matter sector due to the existence of a plethora of potential Lagrangians for ordinary matter and a possible need for a complicated Lagrangian in order to effectively reproduce the dark sector phenomenology~\cite{Capoz}.}. It must be noted that with the addition of dark matter to our model, one must consider the action of the unified gravitational field and of the total matter\footnote{One can clearly see that $S\neq S_O +S_D$.}, which shall be
\begin{equation}
S=\int   \bigg( \frac{1}{16\pi} R + \mathcal{L}_M\bigg)  \sqrt{-g} \, d^4x
\end{equation}
where $\mathcal{L_M}$ describes the unified matter fields, $g=\det\tensor{g}{_{ij}}$ and the total matter is $M=M_O+M_D$. By~varying this action and by following the standard procedure of deriving the field equations we obtain the following unified Einstein equations:
\begin{equation}\label{unEinR}
\tensor{R}{_{ab}}+\Lambda \tensor{g}{_{ab}}-\frac{1}{2}R\tensor{g}{_{ab}}=8\pi\tensor{T}{_{ab}}
\end{equation}
where $\Lambda$ is the cosmological constant and $\tensor{T}{_{ab}}$ is the unified energy-momentum tensor. Using~Equations~(\ref{assumption}), (\ref{riccicurvtensor}) and (\ref{riccicurvscalar}) in (\ref{unEinR}), we obtain the Einstein equation in the following modified form given below\footnote{We assume $c = 1$.}:
\begin{equation}\label{RiemEinstein}
\begin{split}
\big(\tensor{R}{^{(O)}_{ab}}+\Lambda \tensor{g}{^{(O)}_{ab}}-\frac{1}{2}R^{(O)}\tensor{g}{^{ (O)}_{ab}}\big)&+\big(\tensor{R}{^{(D)}_{ab}}+\Lambda\tensor{g}{^{(D)}_{ab}}-\frac{1}{2}R^{(D)}\tensor{g}{^{(D)}_{ab}}\big)
\\ &+\frac{1}{2}\big( 2\tensor{\xi}{_{ab}}-\kappa\tensor{g}{_{ab}} -R^{(O)}\tensor{g}{^{(D)}_{ab}}-R^{(D)}\tensor{g}{^{(O)}_{ab}}\big)= 8\pi G \tensor{T}{_{ab}}   
\end{split}
\end{equation}
where,
\begin{equation}\label{OrdEin}
\tensor{R}{^{(O)}_{ab}}+\Lambda \tensor{g}{^{(O)}_{ab}}-\frac{1}{2}R^{(O)}\tensor{g}{^{ (O)}_{ab}}=8\pi G\tensor{T}{^{(O)}_{ab}}
\end{equation}
\begin{equation}\label{DarkEin}
\tensor{R}{^{(D)}_{ab}}+\Lambda \tensor{g}{^{(D)}_{ab}}-\frac{1}{2}R^{(D)}\tensor{g}{^{ (D)}_{ab}}=8\pi G\tensor{T}{^{(D)}_{ab}}
\end{equation}
and
\begin{equation}\label{intEin}
2\tensor{\xi}{_{ab}}-\kappa \tensor{g}{_{ab}}-R^{(O)}\tensor{g}{^{(D)}_{ab}}-R^{(D)}\tensor{g}{^{(O)}_{ab}} =16\pi G\tensor{\tau}{_{ab}}
\end{equation}
One can clearly observe that Equation~(\ref{OrdEin}) is the Einstein equation for ordinary matter, with~$\tensor{T}{^{(O)}_{ab}}$ being the energy-momentum tensor of ordinary matter. The~corresponding relation for dark matter is also true as can be seen from Equation~(\ref{DarkEin}), where $\tensor{T}{^{(D)}_{ab}}$ plays the role of dark matter's energy-momentum~tensor.

On the other hand, Equation~(\ref{intEin}) is a relation containing terms that hint towards gravitational interaction between ordinary and dark matter. The~tensorial object $\tensor{\tau}{_{ab}}$ acts as the interaction part of the unified energy-momentum tensor since
\begin{equation}\label{totenermom}
\tensor{T}{_{ab}}=\tensor{T}{^{(O)}_{ab}}+\tensor{T}{^{(D)}_{ab}}+\tensor{\tau}{_{ab}}
\end{equation}

As we shall see, the~Einstein Equations%The "Einstein field equations" are commonly referred to as "Einstein equations" (in plural as they are in reality many equations presented in a "compact" tensorial form).
~(\ref{RiemEinstein}) yield interesting results pertaining to the gravitational interaction of dark and ordinary matter. By~taking the trace of Equation~(\ref{RiemEinstein}) we obtain the following~result:
\begin{equation}\label{sigmaR}
\kappa=4\Lambda-8\pi T-R^{(O)}-R^{(D)}
\end{equation}
where $T$ is the trace of the unified energy-momentum~tensor. 

\subsection{Raychaudhuri~Equation}\label{parC}
The Raychaudhuri equation which constitutes an extension of the geodesic deviation equation, plays a significant role in relativity theory and cosmology due to its connection with singularities~\mbox{\cite{Ray, Hawk, Sengupta, Alexiou}}. It governs the behavior and evolution of a family of test particles moving in world lines with given certain variables ($\theta, \sigma, \omega$), where $\theta$ measures the rate of change of the cross-sectional area enclosing a family of geodesics\footnote{The same apply to any curves in general.}, the~shear $\sigma$ measures anisotropy, and~$\omega$ shows a rotation. The~form of the equation depends on the geometry of the spacetime; if a geodesic motion of a family of test particles converges, the~expansion $\theta$ is negative, and~diverges when $\theta$ is~positive.

We shall henceforth consider that both ordinary and dark matter can be considered as perfect fluids, each individually and together as a unified matter perfect fluid comprising the sum of ordinary and dark matter perfect fluids~\cite{Moreshi, Peebles, Alexandre}. The~stress-energy tensor for the perfect fluid distribution of the unified matter (ordinary and dark) with pressure $p$ and energy density $\rho$ falling under its own gravity with 4-velocity $u^a$ is then known to be:
\begin{equation}\label{totenermomtherm}
\tensor{T}{_{ab}}=(\rho + p)u_au_b+p\tensor{g}{_{ab}}
\end{equation}
However, the~unified fluid is composed of two-perfect fluids~\cite{Kuzmichev,Rodolfo} the stress-energy tensor and shall therefore assume the following form~\cite{Ferrando,Oliveira}:
\begin{equation}\label{totenermomtherm2}
\tensor{T}{_{ab}}=(\rho^{(O)}+ p^{(O)})u^{(O)}_au^{(O)}_b+p^{(O)}\tensor{g}{_{ab}}+(\rho^{(D)} + p^{(D)})u^{(D)}_au^{(D)}_b+p^{(D)}\tensor{g}{_{ab}}
\end{equation}
where $p^{(O)}$, $p^{(D)}$ is the pressure of the ordinary and dark matter fluids with energy density $\rho^{(O)}$ and $\rho^{(D)}$ and 4-velocity $u^{(O)}_a$ and $u^{(D)}_a$, respectively.

Using our assumption (\ref{assumption}) Equation~(\ref{totenermomtherm2}) becomes:
\begin{equation}\label{totenermomtherm3}
\begin{split}
\tensor{T}{_{ab}}=(\rho^{(O)}+ p^{(O)})u^{(O)}_au^{(O)}_b&+p^{(O)}\tensor{g}{^{(O)}_{ab}}+(\rho^{(D)} + p^{(D)})u^{(D)}_au^{(D)}_b\\
&+p^{(D)}\tensor{g}{^{(D)}_{ab}}+p^{(O)}\tensor{g}{^{(D)}_{ab}}+p^{(D)}\tensor{g}{^{(O)}_{ab}}      
\end{split}
\end{equation}
One can therefore see that for a perfect fluid and in accordance with Equation~(\ref{totenermom})
\begin{eqnarray}
\tensor{T}{^{(O)}_{ab}}=(\rho^{(O)}+p^{(O)})u^{(O)}_au^{(O)}_b+p^{(O)}\tensor{g}{^{(O)}_{ab}}
\label{partenermomtherm1}\\ \tensor{T}{^{(D)}_{ab}}=(\rho^{(D)} + p^{(D)})u^{(D)}_au^{(D)}_b+p^{(D)}\tensor{g}{^{(D)}_{ab}}
\label{partenermomtherm2}\\ \tensor{\tau}{_{ab}}=p^{(O)}\tensor{g}{^{(D)}_{ab}}+p^{(D)}\tensor{g}{^{(O)}_{ab}}\label{partenermomtherm3}
\end{eqnarray}

We further define the following four operators acting on a vector, e.g.,~$u^a$ (or similarly on a tensor) of $(M,g)$, as~follows:
\begin{equation}\label{slash}
\tensor{u}{^a_{|k}}=\frac{\partial u^a}{\partial x^k} + u^b \tensor{\Gamma}{^{(O)}^a_b_k}
\end{equation}
is the covariant derivative with respect to the ordinary matter subspace, in~other words, the~covariant derivative of a space comprised of purely ordinary matter\footnote{This is true only if the vector (or tensor) acted upon belongs to the space with metric $g^{(O)}$.}.

Similarly,
\begin{equation}
\tensor{u}{^a_{||k}}=\frac{\partial u^a}{\partial x^k} + u^b \tensor{\Gamma}{^{(D)}^a_b_k}
\end{equation}
is the covariant derivative w.r.t. the dark matter subspace\footnote{This is true only if the vector (or tensor) acted upon belongs to the space with metric $g^{(D)}$.}.
\begin{equation}
\tensor{u}{^a_{\textbackslash k}}=\frac{\partial u^a}{\partial x^k} - u^b \tensor{\gamma}{^a_b_k}
\end{equation}
will be an operation that resembles a covariant derivative but uses the interaction symbols, $\tensor{\gamma}{^a_b_c}$\footnote{This operation does not constitute a covariant derivative as the symbols $\tensor{\gamma}{^a_b_c}$ are not proper Christoffel symbols and there is no corresponding geometric space.}. Finally, the~unified space covariant derivative is given by\footnote{One must be careful that only $\tensor{u}{^a_{;k}}$ is the covariant derivative of $u^a \in (M,g)$; all other operations defined before represent arbitrary operators in the framework of the unified space and can only be treated otherwise if we restrict our study in the corresponding subspaces.}:
\begin{equation}
\tensor{u}{^a_{;k}}=\frac{\partial u^a}{\partial x^k} + u^b \tensor{\Gamma}{^a_b_k}= \tensor{u}{^a_{|k}}+\tensor{u}{^a_{||k}}-\tensor{u}{^a_{\textbackslash k}}
\end{equation}

Using this covariant derivative in the rest system of a perfect matter fluid (\ref{totenermomtherm}) the conservation of the energy-momentum tensor is considered \begin{equation}\label{conserv}
\tensor{T}{^{ab}_{;b}}=0
\end{equation}

Assuming that the velocity $u$ corresponding to the total matter fluid (ordinary and dark) is given by $u=u^{(O)}+u^{(D)}$, the~Raychaudhuri equations shall be~\cite{Ray}:
\begin{equation}\label{Rayr}
\dot{\theta}=-2(\sigma^2-\omega^2)-\frac{1}{3}\theta^2 -\tensor{R}{_{ab}}u^au^b+\tensor{\dot{u}}{^a_{;a}}
\end{equation}
where $2\sigma^2=\tensor{\sigma}{_{ab}}\tensor{\sigma}{^a^b}$, $2\omega^2=\tensor{\omega}{_{ab}}\tensor{\omega}{^a^b}$. 

The vorticity tensor, $\tensor{\omega}{_{ab}}=\tensor{u}{_{[a;b]}}-\tensor{\dot{u}}{_{[a}}\tensor{u}{_{b]}}$, is given by
\begin{equation}\label{omega}
\tensor{\omega}{_{ab}}=\tensor{\omega}{^{(O)}_{ab}}+\tensor{\omega}{^{(D)}_{ab}}-\tensor{w}{_{ab}}
\end{equation}
where
\begin{eqnarray}
&\tensor{\omega}{^{(O)}_{ab}}=\tensor{u}{^{(O)}_{[a|b]}}-\tensor{\dot{u}}{^{(O)}_{[a}}\tensor{u}{^{(O)}_{b]}}\\
&\tensor{\omega}{^{(D)}_{ab}}=\tensor{u}{^{(D)}_{[a||b]}}-\dot{u}^{(D)}\tensor{}{_{[a}}\tensor{u}{^{(D)}_{b]}}
\end{eqnarray}
are the vorticity tensors of ordinary and dark matter respectively, and~\begin{equation}
\tensor{w}{_{ab}}=\tensor{u}{_{[a\textbackslash b]}}-\tensor{u}{^{(O)}_{[a||b]}}-\tensor{u}{^{(D)}_{[a|b]}}+ \dot{u}^{(O)}\tensor{}{_{[a}}\tensor{u}{^{(D)}_{b]}}+\dot{u}^{(D)}\tensor{}{_{[a}}\tensor{u}{^{(O)}_{b]}}
\end{equation}
is the interaction part of the vorticity~tensor.

The unified expansion scalar, $\theta=\tensor{u}{^k_{;k}}$ assumes the form
\begin{equation}\label{theta}
\theta=\theta^{(O)}+\theta^{(D)}-\vartheta
\end{equation}
where
\begin{eqnarray}
\theta^{(O)}=\tensor{u}{^k_{|k}}\\
\theta^{(D)}=\tensor{u}{^k_{||k}}
\end{eqnarray}
are the expansion scalars of ordinary and dark matter and
\begin{eqnarray}
\vartheta= \tensor{u}{^{(O)}^k_{||k}}+\tensor{u}{^{(D)}^k_{|k}}-\tensor{u}{^k_{\textbackslash k}}
\end{eqnarray}
is the interaction part of the expansion scalar.

The projection tensor, $\tensor{h}{_{ab}}=\tensor{g}{_{ab}}-u_au_b$, has the following form:
\begin{equation}
\tensor{h}{_{ab}}=\tensor{h}{^{(O)}_{ab}}+\tensor{h}{^{(D)}_{ab}}-\tensor{u}{^{(O)}_{(a}}\tensor{u}{^{(D)}_{b)}}
\end{equation}
where
\begin{eqnarray}
\tensor{h}{^{(O)}_{ab}}=\tensor{g}{^{(O)}_{ab}}-u^{(O)}_au^{(O)}_b,\\
\tensor{h}{^{(D)}_{ab}}=\tensor{g}{^{(D)}_{ab}}-u^{(D)}_au^{(D)}_b
\end{eqnarray}
are the ordinary and dark matter projection~tensors.

The unified shear tensor, $\tensor{\sigma}{_{ab}}=\tensor{u}{_{(a;b)}}-\frac{1}{3}\theta\tensor{h}{_{ab}}-\tensor{\dot{u}}{_{(a}}\tensor{u}{_{b)}}$, shall be given by
\begin{equation}\label{shear}
\tensor{\sigma}{_{ab}}=\sigma^{(O)}\tensor{}{_{ab}}+\sigma^{(D)}\tensor{}{_{ab}}-\tensor{s}{_{ab}}
\end{equation}
where
\begin{eqnarray}
&\sigma^{(O)}\tensor{}{_{ab}}=u^{(O)}\tensor{}{_{(a|b)}}-\frac{1}{3}\theta^{(O)}h^{(O)}\tensor{}{_{ab}}-\dot{u}^{(O)}\tensor{}{_{(a}}u^{(O)}\tensor{}{_{b)}}\\
&\sigma^{(D)}\tensor{}{_{ab}}=u^{(D)}\tensor{}{_{(a||b)}}-\frac{1}{3}\theta^{(D)}h^{(D)}\tensor{}{_{ab}}-\dot{u}^{(D)}\tensor{}{_{(a}}u^{(D)}\tensor{}{_{b)}}
\end{eqnarray}
are the shear tensors of ordinary matter and dark matter fluid respectively, and~\begin{equation}
\begin{split}
\tensor{s}{_{ab}}&=\tensor{u}{_{(a\textbackslash b)}}-u^{(O)}\tensor{}{_{(a||b)}}-u^{(D)}\tensor{}{_{(a|b)}}+
\dot{u}^{(O)}\tensor{}{_{(a}}u^{(D)}\tensor{}{_{b)}}+\dot{u}^{(D)}\tensor{}{_{(a}}u^{(O)}\tensor{}{_{b)}}+\\
&+ \frac{1}{3}\bigg(\theta^{(O)}(h^{(D)}\tensor{}{_{ab}}-u^{(O)}\tensor{}{_{(a}}u^{(D)}\tensor{}{_{b)}})+\theta^{(D)}(h^{(O)}\tensor{}{_{ab}}-u^{(O)}\tensor{}{_{(a}}u^{(D)}\tensor{}{_{b)}})-\vartheta\tensor{h}{_{ab}}\bigg)
\end{split}
\end{equation}
is the interaction part of the shear tensor. Finally, using the Einstein Equation~(\ref{RiemEinstein}) and in conjunction with relations (\ref{partenermomtherm1})--(\ref{partenermomtherm3}) we can find that, apart from the unified Raychaudhuri scalar, $\tensor{R}{_{ab}}u^au^b=4\pi (\rho+3p-\frac{\Lambda}{4\pi})$, which includes ordinary and dark matter as well as their interaction, the~following are also true:
\begin{equation}
\tensor{R}{^{(O)}_{ab}}u^{(O)a} u^{(O)b}=4\pi \bigg(\rho^{(O)}+3p^{(O)}-\frac{\Lambda}{4\pi}\bigg)
\end{equation}
is the Raychaudhuri scalar of ordinary matter and,
\begin{equation}
\tensor{R}{^{(D)}_{ab}}u^{(D)a} u^{(D)b}=4\pi \bigg(\rho^{(D)}+3p^{(D)}-\frac{\Lambda}{4\pi}\bigg)
\end{equation}
is the dark matter Raychaudhuri~scalar. 

By vitrue of relations (\ref{Rayr}), (\ref{omega}), (\ref{theta}) and (\ref{shear}) the expansion $\theta$, the~shear $\sigma$ and the vorticity $\omega$, which~in our model include contributions from dark matter and its interaction with ordinary matter, extend the Raychaudhuri equation which now takes into account the existence of extra mass in the form of dark matter. Since the Raychaudhuri equation plays a dominant role in the evolution of the universe, Equation~(\ref{Rayr}) gives rise to a potential need to differentiate the existing considerations of~singularities. 

\textbf{\subsection{Conformal Dark FLRW-Metric~Structure}\label{parapp}}

In this section we consider an application of our model in cosmology. In~particular, we shall use a FLRW %define
metric structure for the ordinary matter sector and we will assume a conformal relation for the unified spacetime containing both ordinary and dark matter. Using a conformal factor $F(R^{(O)})$ for the unified metric, we shall derive modified Friedmann equations and the continuity equation for a model which is described by $F(R)$ gravity~\cite{Sotiriou,Antonio}. We have chosen such an assumption for the metric of the unified space because studies in the conformal structure in the fields of General Relativity and Cosmology have given rise to viable theories related to dark matter and dark energy (e.g., \cite{Capoz2, Nojiri, Tupper, Hindawi, Felice, Sotiriou, Antonio}) and this framework has been seen to play a significant role as the angles in the light-cone structure are preserved~\cite{Capoz2}. Indeed, we shall further choose a particular class of $F(R^{(O)})$ \cite{Nojiri} and we shall see that the metric of the dark matter sector is also conformal with the ordinary matter FLRW metric structure; namely we assume that
\begin{equation}\label{F(R)}
F(R^{(O)})=R^{(O)}+f(R^{(O)})
\end{equation}
where $F(R^{(O)})$ and $f(R^{(O)})$ are functions of the Ricci scalar of the ordinary matter sector $R^{(O)}$. The~function $F(R^{(O)})$ is the conformal factor of the unified metric and $f(R^{(O)})$ serves as the conformal factor of the dark metric. This is seen as a direct consequence of the assumption (\ref{assumption}) as
\begin{equation}
\tensor{g}{_{ab}} =  F'(R^{(O)})\tensor{g}{^{(O)}_{ab}} = (1+f'(R^{(O)}))\tensor{g}{^{(O)}_{ab}} = \tensor{g}{^{(O)}_{ab}} + f'(R^{(O)})\tensor{g}{^{(O)}_{ab}}
\end{equation}
where $F'(R^{(O)})=\frac{dF(R^{(O)})}{dR^{(O)}}$ and  $f'(R^{(O)})=\frac{df(R^{(O)})}{dR^{(O)}}$.

Alternatively one could start from the assumption that the dark metric is conformal with the ordinary FLRW metric with conformal factor $f(R^{(O)})$ \cite{Capoz2}; namely
\begin{equation}
\tensor{g}{^{(D)}_{ab}} =  f'(R^{(O)})\tensor{g}{^{(O)}_{ab}}
\end{equation}
and using the assumption (\ref{assumption}) we find that
\begin{equation}
\tensor{g}{_{ab}}=\tensor{g}{^{(O)}_{ab}}+\tensor{g}{^{(D)}_{ab}} =  (1+f'(R^{(O)}))\tensor{g}{^{(O)}_{ab}}
\end{equation}

Therefore the unified metric is also conformal with the ordinary FLRW metric if we assume a conformal factor $F(R^{(O)})$ of the form given in (\ref{F(R)}).

Using this previous assumption in conjunction with relations (\ref{Christoff}), (\ref{riccicurvtensor}), (\ref{riccicurvscalar}) and (\ref{totenermom}) we can calculate the Christoffel symbols, the~Ricci curvature as well as the energy momentum tensor\footnote{The velocities are also assumed to follow a conformal relation~\cite{Tupper}.}. From~the Einstein Equation~(\ref{RiemEinstein}) we then derive the following modified Friedmann equations
\begin{equation}\label{F1}
\bigg( \frac{\dot{a}}{a} \bigg)^2=\frac{8\pi G  \rho }{3} F'-\frac{\kappa}{a^2} + \frac{1}{6}\Phi_1(t)
\end{equation}
\begin{equation}\label{F2}
\frac{\ddot{a}}{a} =-\frac{4\pi G  (\rho + 3p)}{3} F' + \frac{1}{6}\Phi_2(t)
\end{equation}
where $a$ is the scale factor, $\kappa=\pm 1, 0$, $\Phi_1(t)=\phi^t\phi_t+\phi^t_{|t}+\phi_{t|t}-\frac{(\phi_t)^2}{2}$, $\Phi_2(t)=\phi^t\phi_t+\phi^t_{|t}-2\phi_{t|t}+(\phi_t)^2$, $\phi_t=\partial_t[\ln{[F']}]$, $\rho=\rho^{(O)}+\rho^{(D)}$, $p=p^{(O)}+p^{(D)}$ and the operator ``$_|$'' has been defined in (\ref{slash}). 

The idea of inflation can be incorporated into the model of Friedmann equations by taking into account the Planck mass $m_{Pl}=(8\pi G)^{-1/2}$ \cite{Carroll,Liddle}. We can then write
\begin{equation}\label{F3}
H^2=\frac{\rho }{3m_{Pl}} F'-\frac{\kappa}{a^2} + \frac{1}{6}\Phi_1(t)
\end{equation}
\begin{equation}\label{F4}
\frac{\ddot{a}}{a} =-\frac{(\rho + 3p)}{6m_{Pl}} F' + \frac{1}{6}\Phi_2(t)
\end{equation}
where $H=\dot{a}/a$ is the Hubble parameter. By~virtue of the above relations we can see that the additional terms $\Phi_i(t)$, $i=1,2$ constitute an extension of the Friedmann equations and may contribute to a differentiation to the accelerated expansion of the universe which is incorporated in the observations. In~a spacetime where dark matter is absent, therefore $f'=0$ or equivalently $F'=1$, it can be easily verified that $\Phi_i(t)$ also vanish and thus, we recover the usual form of the Friedmann equations for a FLRW ordinary matter~space.

Taking into account that the energy momentum tensor of the ordinary sector is conserved, $\tensor{T}{^{(O)\mu} _{0 | \mu}}=0$, we can see that the energy momentum tensor of the unified two component cosmological fluid is also conserved and using the fact that $\tensor{T}{^\mu _{0 ; \mu}}=0$ we derive the following continuity equation
\begin{equation}
\dot{\rho}+3(1+w)\rho H=0
\end{equation}
where we assume that the total pressure $p=w\rho$. This result is also compatible with the general conservation relation (\ref{conserv}).

\section{Gravity on the Sasaki Tangent~Bundle}\label{section3}
For the study of the tangent bundle we use a metric, $g$, on~a Riemannian manifold $(M,g)$, along~with its Levi-Civita connection, to~construct a new natural Riemannian metric on the tangent bundle $TM$, called the {Sasaki metric}. The~geometry of tangent bundles in general and particularly the Sasaki metric find a lot of applications in physics, especially in the study of gravity. This consideration extends the limits of conventional general relativity for a modified gravity approach, and~provides the gravitational field with extra degrees of freedom. In~particular, a~Sasaki extension of spacetime constitutes the minimum metric generalization of spacetime in the framework of a tangent bundle of a four-dimensional~spacetime.

\subsection{Deviation of Geodesics of a Sasaki~Spacetime}\label{parD}
In this section, we shall derive the deviation of geodesics of a Sasaki spacetime with metric~\cite{Sas-58}
\begin{equation}\label{msas}
d\sigma^2=g_{\alpha\beta}(x)dx^\alpha dx^\beta + g_{\alpha\beta}(x)D y^\alpha D y^\beta
\end{equation}
where,
\begin{equation}\label{Delt}
Dy^\alpha = dy^\alpha + \tensor{N}{^\alpha_\beta}(x,y) dx^\beta 
\end{equation}
$\tensor{N}{^\mu_\alpha}(x,y)$ represents the ``pre-Finsler non-linear connection'', $g_{\alpha\beta}(x)$ is the Riemann metric tensor of the n-dimensional differentiable manifold (here n = 4) and both the horizontal and vertical part of the total metric (\ref{msas}) of the 2n-dimensional tangent bundle. $\alpha,\beta \in \{1,2,...,n=4\}$.

Furthermore, for~the purpose of our study, the~non-linear connection is assumed to be
\begin{equation}\label{mnonlin}
\tensor{N}{^\mu_\alpha}=\tensor{\Gamma}{^\mu_{\alpha\kappa}}y^\kappa
\end{equation}

The metric (\ref{msas}) may, then be rewritten using the fundamental covariant metric tensor, $\tensor{G}{_{ij}}$ as
\begin{equation}\label{fundmetr}
d\sigma^2=\tensor{G}{_{ij}}dx^i dx^j
\end{equation}
where $i,j \in \{1,2,...,2n=8\}$ and
\begin{equation}
\tensor{G}{_{\alpha\beta}}=\tensor{g}{_{\alpha\beta}}+\tensor{g}{_{\mu\nu}}\tensor{\Gamma}{^\mu_{\rho\alpha}}\tensor{\Gamma}{^\nu_{\kappa\beta}}y^\rho y^\kappa
\end{equation}
\begin{equation}
\tensor{G}{_{\alpha(n+\beta)}}=\tensor{\Gamma}{_\mu_{\alpha\beta}}y^\mu 
\end{equation}
and
\begin{equation}
\tensor{G}{_{(n+\alpha)(n+\beta)}}=\tensor{g}{_{\alpha\beta}}
\end{equation}
where $\tensor{\Gamma}{_\mu_{\alpha\beta}}$ and $\tensor{\Gamma}{^\mu_{\rho\alpha}}$ are Christoffel's symbols of the first and second kind of $M^n$ respectively.

If $\bar{\Gamma}\tensor{}{_{ijk}}$ and $\bar{\Gamma}\tensor{}{^i_{jk}}$ are Christoffel's symbols of first and second kind, respectively, that correspond to the Sasaki tangent bundle $T(M^n)$ as calculated using the fundamental metric, then the geodesic equation is known to be
\begin{equation}\label{sasfundgeod}
\frac{d^2x^i}{d\sigma^2}+\tensor{\bar{\Gamma}}{^i_j_k}\frac{dx^j}{d\sigma}\frac{dx^k}{d\sigma}=0
\end{equation}

While it is usually more useful to express this equation in terms of quantities of $M^n$, for~the present we shall content ourselves with using relation (\ref{sasfundgeod}) instead, in~order to avoid perplexing our~equations. 

If $x^i_s(t)$ is a two-parameter geodesic family with $t$ being the affine parameter and $s$ the selector parameter then we shall denote the tangent vectors as $V^i=\partial_t x^i$, $W^i=\partial_s x^i$ so that $\partial_sV^i=\partial_t W^i$. Moreover, let $X^i(x^k)$ be a vector field defined over a region of the subspace of $T(M^n)$ defined by the net of $x^i_s(t)$. Then we define the $\delta-$derivatives as:
\begin{equation}
\frac{\delta X^i}{\delta t}=\tensor{X}{^i_{;h}}V^h=\Big(\frac{\partial X^i}{\partial x^h}+\tensor{\bar{\Gamma}}{^i_h_k} X^k\Big)V^h
\end{equation}
and
\begin{equation}
\frac{\delta X^i}{\delta s}=\tensor{X}{^i_{;h}}W^h=\Big(\frac{\partial X^i}{\partial x^h}+\tensor{\bar{\Gamma}}{^i_h_k} X^k\Big)W^h
\end{equation}

Using the previous result in conjunction with the fact that the Christoffel's symbols are symmetric w.r.t. their suffixes, we get that
\begin{equation}
\frac{\delta V^i}{\delta s}=\frac{\partial^2 x^i}{\partial t \, \partial s} +\tensor{\bar{\Gamma}}{^i_h_k} V^kW^h= \frac{\delta W^i}{\delta t}
\end{equation}

Following the standard procedure of deriving the deviation equations~\cite{Rund}, given the previous relations, we find that
\begin{equation}
\frac{\delta^2 W^i}{\delta t^2}=\frac{\delta^2 V^i}{\delta s \, \delta t}=\frac{\delta^2 V^i}{\delta t \, \delta s}- \tensor{K}{^i_{jhk}}V^jV^hW^k
\end{equation}

Finally, using Equation~(\ref{sasfundgeod}) we find the following geodesic deviation equation
\begin{equation}\label{sasadev}
\frac{\delta^2 W^i}{\delta t^2}=-\tensor{K}{^i_{jhk}}V^jV^hW^k
\end{equation}
where $\tensor{K}{^i_{jhk}}$ is the curvature tensor of the tangent bundle. An~abstract form of the curvature tensor is given by~\cite{French,German}. All the components of the curvature tensor $\tensor{K}{^i_{jhk}}$ are given explicitly in Appendix \ref{App}.

\subsection{Dark Gravity on the Tangent~Bundle}\label{parE}
Experimental research~\cite{Mukhanov, Nature, Question, Brazil, Group} suggests that a theory of dark gravity with extra dimensions may be necessary in order to effectively describe the total mass distribution in the universe. A~first step towards such a geometric gravitational theory could be obtained by retaining our assumption (\ref{assumption}) and expanding our Riemannian framework from Section~\ref{section2} on the tangent bundle using an underlying Sasaki structure (\ref{msas}) for the total space of ordinary and dark~matter. 

First, taking into account relation (\ref{mnonlin}) and using Equation~(\ref{Christoff}) we find that
\begin{equation}\label{nonlin}
\tensor{N}{^\mu_\alpha}=\tensor{N}{^{(O)}^\mu_\alpha}+\tensor{N}{^{(D)}^\mu_\alpha}+\tensor{\nu}{^\mu_\alpha}
\end{equation}
where,
\begin{equation}
\tensor{N}{^{(O)}^\mu_\alpha}=\tensor{\Gamma}{^{(O)}^\mu_{\alpha\kappa}}y^\kappa
\end{equation}
is the ordinary matter non-linear connection,
\begin{equation}
\tensor{N}{^{(D)}^\mu_\alpha}=\tensor{\Gamma}{^{(D)}^\mu_{\alpha\kappa}}y^\kappa
\end{equation}
is dark matter's non-linear connection, and~\begin{equation}
\tensor{\nu}{^\mu_\alpha}=\tensor{\gamma}{^\mu_{\alpha\kappa}}y^\kappa
\end{equation}
is the interaction part of the non-linear connection. $\tensor{\nu}{^\mu_\alpha}$ acts as a correlation between the non-linear connections of ordinary and dark matter. Thus, in~the framework of a Sasaki spacetime, $\tensor{\nu}{^\mu_\alpha}$ plays the fundamental role of interconnecting the ordinary and dark matter sectors within the line-element. Such a connection between ordinary and dark matter is absent from the corresponding line-element of the Riemannian spacetime. It can therefore be concluded that a Sasaki spacetime involves a stronger interaction between the two component matter sectors which influences even the notion of arc-length as we shall also see~below.

In virtue of relations (\ref{Delt}) and (\ref{nonlin}), we obtain
\begin{equation}\label{covdifferential}
Dy^\mu =  D^{(O)}y^\mu+ D^{(D)}y^\mu - \delta y^\mu
\end{equation}
where,
\begin{equation}
D^{(O)}y^\mu = dy^\mu + \tensor{N}{^{(O)}^\mu_\alpha}(x,y) dx^\alpha
\end{equation}
\begin{equation}
D^{(D)}y^\mu = dy^\mu + \tensor{N}{^{(D)}^\mu_\alpha}(x,y) dx^\alpha 
\end{equation}
and
\begin{equation}
\delta y^\mu = dy^\mu - \tensor{\nu}{^\mu_\alpha}(x,y) dx^\alpha
\end{equation}

In a Sasaki spacetime , the~non-linear connection influences the gravitational potential giving rise to extra degrees of freedom in the internal structure of the spacetime in the form of $y$-dependence. As~evidenced by relations (\ref{assumption}),  (\ref{msas}), (\ref{Delt}) and (\ref{covdifferential}), with~the addition of dark matter on the tangent bundle of the spacetime , three types of line-elements can be presented:
\begin{equation}
d\sigma^2=d\sigma_O^2+d\sigma_D^2+d\sigma_I^2
\end{equation}
where
\begin{equation}
d\sigma_O^2= \tensor{g}{^{(O)}_{\alpha \beta}}dx^\alpha dx^\beta + \tensor{g}{^{(O)}_{\alpha \beta}}D^{(O)}y^\alpha D^{(O)}y^\beta
\end{equation}
is the line-element of the Sasaki tangent bundle of ordinary matter,
\begin{equation}
d\sigma_D^2= \tensor{g}{^{(D)}_{\alpha \beta}}dx^\alpha dx^\beta + \tensor{g}{^{(D)}_{\alpha \beta}}D^{(D)}y^\alpha D^{(D)}y^\beta
\end{equation}
is the line-element of the Sasaki tangent bundle of dark matter, and~\begin{equation}
d\sigma_I^2=\tensor{g}{_{\alpha \beta}}Dy^\alpha Dy^\beta -\tensor{g}{^{(O)}_{\alpha \beta}}D^{(O)}y^\alpha D^{(O)}y^\beta-\tensor{g}{^{(D)}_{\alpha \beta}}D^{(D)}y^\alpha D^{(D)}y^\beta
\end{equation}
is the interaction term of the~line-element. 

It can be seen that the non-linear connection, $N(x,y)$, plays a fundamental role in this approach, since it facilitates the introduction of three differentials (\ref{covdifferential}) that produce extra interaction terms between ordinary and dark matter. In~the base manifold $(M,g)$ of Section~\ref{section2}, the~introduction of dark matter extends the arc-length of the spacetime of ordinary matter in an additive way since one could write relation (\ref{lineel}) as
\begin{equation}
ds^2=ds_O^2+ds_D^2
\end{equation}
where
\begin{equation}
ds_O^2= \tensor{g}{^{(O)}_{\alpha \beta}}dx^\alpha dx^\beta
\end{equation}
is the line-element of an ordinary matter spacetime, and~\begin{equation}
ds_D^2= \tensor{g}{^{(D)}_{\alpha \beta}}dx^\alpha dx^\beta
\end{equation}
is the line-element of a dark matter spacetime. Such ``trivial'' extension is not possible in the Sasaki tangent bundle because of the non-linear connection introducing extra interaction that influences even the arc-length~itself.

We have seen that the geodesics are provided by the condensed relation (\ref{sasfundgeod}) which is known to be expanded in the following two equations~\cite{Sas-58}:
\begin{equation}\label{hsasageod}
\frac{d^2x^\mu}{d\sigma^2}+\tensor{\Gamma}{^\mu_{\nu \rho}}\frac{dx^\nu }{d\sigma}\frac{dx^\rho}{d\sigma}=\tensor{R}{^\mu_{\nu \alpha \beta}}\frac{dx^\nu }{d\sigma}y^\alpha \frac{Dy^\beta}{d\sigma}
\end{equation}
\begin{equation}\label{vsasageod}
\frac{D^2y^\mu}{d\sigma^2}=0
\end{equation}
where $\tensor{\Gamma}{^\mu_{\nu \rho}}$ and $\tensor{R}{^\mu_{\nu \alpha \beta}}$ are the Christoffel symbols and Riemann curvature tensor, respectively, corresponding to the base manifold $(M,g)$ as discussed in Section~\ref{section2}, and~\begin{equation}\label{1covd}
\frac{Dy^\mu}{d\sigma}=\frac{dy^\mu}{d\sigma}+\tensor{\Gamma}{^\mu_{\nu \rho}}y^\nu \frac{dx^\rho}{d\sigma}
\end{equation}
\begin{equation}\label{2covd}
\frac{D^2y^\mu}{d\sigma^2}=\frac{d^2y^\mu}{d\sigma^2}+\frac{d}{d\sigma}\bigg\{\tensor{\Gamma}{^\mu_{\nu \rho}}y^\nu \frac{dx^\rho}{d\sigma}\bigg\}+\tensor{\Gamma}{^\mu_{\alpha \beta}}\frac{Dy^\alpha}{d\sigma}\frac{dx^\beta}{d\sigma}
\end{equation}

The expansion of the condensed geodesic Equation~(\ref{sasfundgeod}) gave rise to two geodesic equations. The~horizontal geodesic (\ref{hsasageod}) can be physically interpreted as the generalization of the corresponding Riemannian geodesic curves on the tangent bundle. The~expanded family of geodesic curves described by (\ref{hsasageod}) shall be further examined below. On~the other hand, the~physical interpretation of the vertical Equation~(\ref{vsasageod}) remains an open question and although we shall continue to provide the vertical equations, we shall limit our present work to the physical study of the horizontal~geodesics.

If we compare the Riemannian setting with the Sasaki tangent bundle, we can see from relation~(\ref{hsasageod}) that the class of curves that have the geodesic property has now been expanded. On~$(TM,G)$ we can distinguish two geodesic families the first of which is obtained by lifting a geodesic of $(M,g)$ on $TM$. In~that case the expanded form of the geodesic equations reduces as~\cite{Sas-58}
\begin{equation}\label{hsasageodM}
\frac{d^2x^\mu}{d\sigma^2}+\tensor{\Gamma}{^\mu_{\nu \rho}}\frac{dx^\nu }{d\sigma}\frac{dx^\rho}{d\sigma}=0
\end{equation}
\begin{equation*}
\frac{Dy^\mu}{d\sigma}=0
\end{equation*}

In this case, with~reference to the base manifold $M$ of $TM$, an~observer  shall continue to perceive a state of rest since Equation~(\ref{hsasageodM}) coincides with their notion of a Riemannian geodesic. We will henceforth call an observer limited to perceiving only a Riemannian manifold occupied by ordinary matter i.e.,~neglecting dark matter and ignoring the existence of any higher-dimensional structure, a~{constrained observer}. In~essence, ``constrained'' observers think that they are experiencing the submanifold $(M^{(O)},g^{(O)})$ of the base manifold, thereby perceiving anything deviating from their notion of a Riemannian structure as due to external forces or effects. Such interpretation is compatible with the idea of an apparent metric as proposed by~\cite{Capoz2}.

The second family of geodesics on the Sasaki tangent bundle is comprised by curves that are not obtained by lifting geodesics of $M$. In~that case Equation~(\ref{hsasageod}) cannot be further reduced. This causes a constrained observer to perceive apparent {pseudo-coupling forces} of the gravitational field with the velocity field of $TM$, preventing them from understanding that curve as a geodesic. The~r.h.s. part of Equation~(\ref{hsasageod}) seemingly disagrees with the notion of a Riemannian geodesic. However, this effect is only due to the observer's inability to perceive $TM$.

By virtue of relations (\ref{covdifferential}) and (\ref{1covd}) we have that:
\begin{equation}\label{dies}
\frac{Dy^\mu}{d\sigma}=\frac{D^{(O)}y^\mu}{d\sigma}+\frac{D^{(D)}y^\mu}{d\sigma}-\frac{\delta y^\mu}{d\sigma}
\end{equation}
where
\begin{equation}
\frac{D^{(O)}y^\mu}{d\sigma}=\frac{dy^\mu}{d\sigma}+\tensor{\Gamma}{^{(O)}^\mu_{\nu \rho }}y^\nu \frac{dx^\rho }{d\sigma}
\end{equation}
\begin{equation}
\frac{D^{(D)}y^\mu}{d\sigma}=\frac{dy^\mu}{d\sigma}+\tensor{\Gamma}{^{(D)}^\mu_{\nu \rho }}y^\nu \frac{dx^\rho }{d\sigma}
\end{equation}
\begin{equation}
\frac{\delta y^\mu}{d\sigma}=\frac{dy^\mu}{d\sigma}-\tensor{\gamma}{^\mu_{\nu \rho }}y^\nu \frac{dx^\rho }{d\sigma}
\end{equation}

Using Equations~(\ref{2covd}) and (\ref{dies}) we get:
\begingroup\makeatletter\def\f@size{8}\check@mathfonts
\def\maketag@@@#1{\hbox{\m@th\normalsize\normalfont#1}}%
\begin{equation}
\frac{D^2y^\mu}{d\sigma^2}=\frac{D^{(O)2}y^\mu}{d\sigma^2}+\frac{D^{(D)2}y^\mu}{d\sigma^2}+\frac{\delta^2 y^\mu}{d\sigma^2}+\frac{D^{(O)}}{d\sigma}\bigg\{\frac{D^{(D)}y^\mu}{d\sigma}-\frac{\delta y^\mu}{d\sigma}\bigg\}+\frac{D^{(D)}}{d\sigma}\bigg\{\frac{D^{(O)}y^\mu}{d\sigma}-\frac{\delta y^\mu}{d\sigma}\bigg\}-\frac{\delta}{d\sigma}\bigg\{\frac{D^{(O)}y^\mu}{d\sigma}+\frac{D^{(D)}y^\mu}{d\sigma}\bigg\}
\end{equation}
\endgroup
where
\begin{equation}
\frac{D^{(O)2}y^\mu}{d\sigma^2} =\frac{d^2y^\mu}{d\sigma^2}+\frac{d}{d\sigma}\bigg\{\tensor{\Gamma}{^{(O)}^\mu_{\nu \rho }}y^\nu \frac{dx^\rho }{d\sigma}\bigg\}+\tensor{\Gamma}{^{(O)}^\mu_{\alpha \beta}}\frac{D^{(O)}y^\alpha}{d\sigma}\frac{dx^\beta}{d\sigma}
\end{equation}
\begin{equation}
\frac{D^{(D)2}y^\mu}{d\sigma^2} =\frac{d^2y^\mu}{d\sigma^2}+\frac{d}{d\sigma}\bigg\{\tensor{\Gamma}{^{(D)}^\mu_{\nu \rho }}y^\nu \frac{dx^\rho }{d\sigma}\bigg\}+\tensor{\Gamma}{^{(D)}^\mu_{\alpha \beta}}\frac{D^{(D)}y^\alpha}{d\sigma}\frac{dx^\beta}{d\sigma}
\end{equation}
\begin{equation}
\frac{\delta^2 y^\mu}{d\sigma^2} =\frac{d^2y^\mu}{d\sigma^2}+\frac{d}{d\sigma}\bigg\{\tensor{\gamma}{^\mu_{\nu \rho }}y^\nu \frac{dx^\rho }{d\sigma}\bigg\}+\tensor{\gamma}{^\mu_{\alpha \beta}}\frac{\delta y^\alpha}{d\sigma}\frac{dx^\beta}{d\sigma}
\end{equation}

From the point of view of an observer limited to the ordinary matter subspace relations (\ref{rgeod}) and~(\ref{testp}) show that the geodesic equations in the Riemannian setting are perturbed when dark matter is taken into account. Let us now assume a test particle moving along geodesics on the tangent bundle in a Sasaki spacetime. In~analogy to the Riemannian case the geodesics in higher dimensions will deviate from the previously thought geodesic motion due to the presence of dark matter and its interaction with ordinary matter. In~this case, using our previous results in conjunction with relations (\ref{Christoff}), (\ref{riemanncurvtensor}) and~(\ref{dies}) in Equations~(\ref{hsasageod}) and (\ref{vsasageod}), the~unified form of the corresponding geodesic equation will be given by the following:
\begingroup\makeatletter\def\f@size{8}\check@mathfonts
\def\maketag@@@#1{\hbox{\m@th\normalsize\normalfont#1}}%
\begin{eqnarray}\label{dhsasageod}
\frac{d^2x^\mu}{d\sigma^2}+\big(\tensor{\Gamma}{^{(O)}^\mu_{\nu \rho }}+\tensor{\Gamma}{^{(D)}^\mu_{\nu \rho }}+\tensor{\gamma}{^\mu_{\nu \rho }}\big)\frac{dx^\nu }{d\sigma}\frac{dx^\rho }{d\sigma}=\big(\tensor{R}{^{(O)}^\mu_{\nu \alpha \beta}}+\tensor{R}{^{(D)}^\mu_{\nu \alpha \beta}}+\tensor{r}{^\mu_{\nu \alpha \beta}}\big)\frac{dx^\nu }{d\sigma}y^a\bigg(\frac{D^{(O)}y^b}{d\sigma}+\frac{D^{(D)}y^b}{d\sigma}-\frac{\delta y^b}{d\sigma}\bigg)\nonumber\\
\end{eqnarray}
\endgroup
and
\begingroup\makeatletter\def\f@size{8.5}\check@mathfonts
\def\maketag@@@#1{\hbox{\m@th\normalsize\normalfont#1}}%
\begin{equation}
\frac{D^{(O)2}y^\mu}{d\sigma^2}+\frac{D^{(D)2}y^\mu}{d\sigma^2}+\frac{\delta^2 y^\mu}{d\sigma^2}=-\frac{D^{(O)}}{d\sigma}\bigg\{\frac{D^{(D)}y^\mu}{d\sigma}-\frac{\delta y^\mu}{d\sigma}\bigg\}-\frac{D^{(D)}}{d\sigma}\bigg\{\frac{D^{(O)}y^\mu}{d\sigma}-\frac{\delta y^\mu}{d\sigma}\bigg\}+\frac{\delta}{d\sigma}\bigg\{\frac{D^{(O)}y^\mu}{d\sigma}+\frac{D^{(D)}y^\mu}{d\sigma}\bigg\} 
\end{equation}
\endgroup

As we have discussed, the~expanded family of geodesics of a Sasaki spacetime includes curves that can be obtained by lifting geodesics of the base manifold. In~this case, from~the perspective of a constrained observer who takes into account only the ordinary matter sector of the Riemannian space $(M,g)$, the~test particle moves along a curve that deviates from their expected Riemannian geodesics due to the presence of {dark pseudo-forces}\footnote{These are the same $F^i_{(D)}$ as in Equation~(\ref{testp}).}. If~however, the~test particle at rest happens to be moving along a Sasaki geodesic that cannot be obtained by such a lift, then the constrained observer will detect, in~addition to the {dark pseudo-forces}, the~{pseudo-coupling forces} given by the r.h.s. part of Equation~(\ref{dhsasageod}). Therefore, in~a Sasaki spacetime a constrained observer shall always observe deviated curves as is the case with a Riemannian spacetime. The~difference between a Riemannian and a Sasaki setting lies in the observation of the {pseudo-coupling forces} which may or may not be detected depending on the~geodesic.

By virtue of relation (\ref{sasacurv}) it can be shown that
\begin{equation}
\tensor{K}{^i_{jhk}}=\tensor{K}{^{(O)}^i_{jhk}}+\tensor{K}{^{(D)}^i_{jhk}}+\tensor{\kappa}{^i_{jhk}}
\end{equation} 
where $\tensor{K}{^{(O)}^i_{jhk}}$, $\tensor{K}{^{(D)}^i_{jhk}}$, are the Sasaki curvature tensors of ordinary and dark matter, respectively, and~$\tensor{\kappa}{^i_{jhk}}=\tensor{K}{^i_{jhk}}-\tensor{K}{^{(O)}^i_{jhk}}-\tensor{K}{^{(D)}^i_{jhk}}$.

Using relation (\ref{sasadev}) with the above result we get
\begin{equation}\label{dsasadev}
\frac{\delta^2 W^i}{\delta t^2}=-(\tensor{K}{^{(O)}^i_{jhk}}+\tensor{K}{^{(D)}^i_{jhk}}+\tensor{\kappa}{^i_{jhk}})V^jV^hW^k
\end{equation}

As we have seen from relation (\ref{sasacurv}), the~curvature tensor, $K$, has been modified in higher dimensions due to the geometry of the Sasaki tangent bundle. Consequently, we can conclude that the gravitational field has also been influenced by the higher-dimensional metric structure of the spacetime, i.e.,~the Sasaki tangent bundle, since gravity is intertwined with the notion of curvature. The~gravitational field has been endowed with extra degrees of freedom and this has been incorporated in the tidal forces (\ref{dsasadev}) by means of the generalized curvature tensor. As~a result, the~deviation of the 4-dimensional free motion of nearby particles (or clusters thereof, e.g.,~galaxies) (\ref{rdev}) can be differentiated from its higher-dimensional counterpart because of the form of the curvature tensor which is an intrinsic geometric property of the spacetime and therefore independent of the relative observer. Therefore, the~resulting {dark tidal forces} manifest both the addition of extra matter and interaction in the form of dark matter as well as the higher-dimensional geometric structure of the spacetime and the subsequent extra degrees of freedom of the gravitational~field.

The previous result is of great importance for cosmology. For~instance, a~positive curvature of a Riemannian spacetime which is connected with converging neighbouring geodesics (Jacobi field) may correspond to a negative curvature on the Sasaki tangent bundle. Hence, taking into account the higher-dimensional structure could potentially reveal that what we initially thought of as convergent is in reality divergent, and~vice-versa, due to extra forces stemming from the geometry of the~spacetime.  

\section{Concluding~Remarks}\label{conclusion}
Motivated by observational results, we laid the foundations for a geometric theory of dark gravity in relation to the ordinary gravitational field in the framework of a ``unified'' space. By~extending the already existing notions of geometry that apply for ordinary matter, we used them in the case of dark gravity. In~particular, in~this work we examined the unified gravitational field of ordinary and dark matter in a Riemannian metric framework deriving the geodesic equations and their deviation, as~well as the Einstein and Raychaudhuri equations. Furthermore, by~extending our geometric methods on the tangent bundle we presented a unified dark and ordinary gravity on a Sasaki~spacetime.

The geometric methods used throughout this paper gave rise to the concept of dark forces which could play a role in offsetting the centrifugal orbital tendency of the galaxies. In~particular, the~gravitational influence of dark matter generated modified geodesic curves that seemingly deviate from the Riemannian geodesic notion due to the presence of {dark pseudo-forces} (\ref{testp}) and (\ref{dhsasageod}) which appear to dominate the deviated geodesic motion. The~extension of the spacetime on the tangent bundle also caused the potential emergence of the {pseudo-coupling forces} (\ref{dhsasageod}) due to the inability of the observer to perceive the higher dimensional structure. Both of these cases are deemed as pseudo-forces because they are the product of a constrained observer. The~third kind of dark forces however, the~{dark tidal forces} (\ref{rdev}) and (\ref{dsasadev}), result from an intrinsic geometric property of the spacetime; the dark curvature. This shows that the gravitational interaction between ordinary and dark matter has been incorporated in the geometry of the spacetime and therefore, {dark tidal forces} exist independent of the~observer.

With the introduction of dark matter to the Riemannian spacetime setting, the~Einstein and Raychaudhuri equations were modified to account for the existence of extra matter and thus, for~extra gravitational interaction. These equations could provide invaluable information concerning the interaction of dark and ordinary matter and hence provide details about the internal structure of the unified space. The~modification of these equations due to dark matter resulted in modified Friedmann Equations~(\ref{F1})--(\ref{F4}) as examined using a conformal metric structure for the dark matter sector. The~study of inflation and the cosmological implications of these relations in the evolution of the universe as well as the application of other models of dark metric structure could constitute the subject of further~research.

Overall, the~existence of dark matter calls into question the concept of geodesic motion. Indeed~the appearance of pseudo-force fields dominating the motion of test particles causes a constrained observer to conclude that the motion is not geodesic and that the ordinary matter spacetime has been externally perturbed by dark matter. Nonetheless, a~closer look at the overall structure of the spacetime reveals that those apparent forces exist only relative to the point of view of a particular observer and reaffirms the geodesic property of the~curves. 

Ultimately, we consider that the use of geometric methods in the formulation of a gravitational theory that includes dark matter could form a theoretical basis upon which observational results could be interpreted. The~contribution of dark energy in the framework of this theoretical effort remains an even greater question for future~research.

\vspace{6pt}
\noindent{\bf Conflicts of interest:} The authors declare no conflict of interest.

\appendix

\section{The Curvature Tensor of a Sasaki Tangent Bundle}\label{App}
Using the Christoffel symbols $\bar{\Gamma}\tensor{}{^i_{jk}}$ as given by S. Sasaki in relation (7.4) in~\cite{Sas-58} and after long calculations, it can be seen that the curvature tensor of a $2n$-dimensional Sasaki tangent bundle $(TM,G)$~is
\begin{equation}\label{sasacurv}
\tensor{K}{^i_{(n+\alpha)(n+\beta)(n+\gamma)}}=0
\end{equation}
\begin{eqnarray*}
	\tensor{K}{^\delta_{(n+\alpha)(n+\beta) \gamma}}&=&\frac{1}{2}\frac{\partial}{\partial x^{(n+\beta)}}[\tensor{R}{^\delta _{\gamma \alpha \lambda}} y^\lambda] +\frac{1}{4} \tensor{R}{^\delta _{\epsilon \beta \mu}} \tensor{R}{^\epsilon _{\gamma \alpha \nu}} y^\mu y^\nu 
\end{eqnarray*}

\begin{eqnarray*}
	\tensor{K}{^{(n+\delta)} _{(n+\alpha)(n+\beta) \gamma}}&=&
	\frac{\partial}{\partial x^{(n+\beta)}}
	[\tensor{\Gamma}{^\delta _{\alpha \gamma}} -\frac{1}{2} \tensor{\Gamma}{^\delta _{\mu \lambda}}\tensor{R}{^\lambda _{\gamma \alpha \nu}}y^\mu y^\nu] +\frac{1}{2} [\tensor{\Gamma}{^\delta _{\beta \epsilon}} -\frac{1}{2} \tensor{\Gamma}{^\delta _{\mu \lambda}}\tensor{R}{^\lambda _{\epsilon \beta \nu}}y^\mu y^\nu]\tensor{R}{^\epsilon _{\gamma \alpha \kappa}}y^\kappa 
\end{eqnarray*}

\begingroup\makeatletter\def\f@size{9.7}\check@mathfonts
\def\maketag@@@#1{\hbox{\m@th\normalsize\normalfont#1}}%
\begin{eqnarray*}
	\tensor{K}{^\delta_{(n+\alpha) \beta \gamma}}&=&\frac{1}{2} \frac{\partial}{\partial x^{\beta}}[\tensor{R}{^\delta _{\gamma \alpha \lambda}} y^\lambda]- \frac{1}{2}\frac{\partial}{\partial x^{\gamma}}[\tensor{R}{^\delta _{\beta \alpha \lambda}} y^\lambda]+\frac{1}{2} [\tensor{\Gamma}{^\delta _{\beta \epsilon}}+\frac{1}{2}(\tensor{R}{^\delta _{\epsilon \kappa \mu}}\tensor{\Gamma}{^\kappa _{\nu \beta}}+\tensor{R}{^\delta _{\beta \kappa \mu}}\tensor{\Gamma}{^\kappa _{\nu \epsilon}})y^\mu y^\nu]\tensor{R}{^\epsilon _{\gamma \alpha \lambda}}y^\lambda\\
	&-& \frac{1}{2} [\tensor{\Gamma}{^\delta _{\gamma \epsilon}}+\frac{1}{2}(\tensor{R}{^\delta _{\epsilon \kappa \mu}}\tensor{\Gamma}{^\kappa _{\nu \gamma}}+\tensor{R}{^\delta _{\gamma \kappa \mu}}\tensor{\Gamma}{^\kappa _{\nu \epsilon}})y^\mu y^\nu]\tensor{R}{^\epsilon _{\beta \alpha \lambda}}y^\lambda+ \frac{1}{2} [\tensor{\Gamma}{^\epsilon _{\alpha \gamma}} -\frac{1}{2}  \tensor{\Gamma}{^\epsilon _{\mu \kappa}}\tensor{R}{^\kappa _{\gamma \alpha \nu}}y^\mu y^\nu ]\tensor{R}{^\delta _{\beta \epsilon \lambda}} y^\lambda\\ 
	&-& \frac{1}{2} [\tensor{\Gamma}{^\epsilon _{\alpha \beta}} -\frac{1}{2}  \tensor{\Gamma}{^\epsilon _{\mu \kappa}}\tensor{R}{^\kappa _{\beta \alpha \nu}}y^\mu y^\nu ]\tensor{R}{^\delta _{\gamma \epsilon \lambda}} y^\lambda
\end{eqnarray*}
\endgroup

\begin{eqnarray*}
	\tensor{K}{^{(n+\delta)}_{(n+\alpha) \beta \gamma}}&=&\frac{\partial}{\partial x^{\beta}}[\tensor{\Gamma}{^\delta _{\alpha \gamma}} -\frac{1}{2} \tensor{\Gamma}{^\delta _{\mu \lambda}}\tensor{R}{^\lambda _{\gamma \alpha \nu}}y^\mu y^\nu]-\frac{\partial}{\partial x^{\gamma}}[\tensor{\Gamma}{^\delta _{\alpha \beta}} -\frac{1}{2} \tensor{\Gamma}{^\delta _{\mu \lambda}}\tensor{R}{^\lambda _{\beta \alpha \nu}}y^\mu y^\nu]\\
	&+&\frac{1}{4}\tensor{R}{^\epsilon _{\gamma \alpha \xi}}y^\xi [(\tensor{R}{^\delta _{\beta \lambda \epsilon}} +\tensor{R}{^\delta _{\epsilon \lambda \beta}} + 2\frac{\partial \tensor{\Gamma}{^\delta _{\beta \epsilon}}}{\partial x^\lambda})y^\lambda + \tensor{\Gamma}{^\delta _{\nu \kappa}}(\tensor{R}{^\kappa _{\epsilon \mu \eta}} \tensor{\Gamma}{^\eta _{\lambda \beta}}+\tensor{R}{^\kappa _{\beta \mu \eta}} \tensor{\Gamma}{^\eta _{\lambda \epsilon}})y^\lambda y^\mu y^\nu]\\
	&+&[\tensor{\Gamma}{^\delta _{\epsilon \beta}} - \frac{1}{2} \tensor{\Gamma}{^\delta _{\mu \kappa}}\tensor{R}{^\kappa _{\beta \epsilon \nu}} y^\mu y^\nu][\tensor{\Gamma}{^\epsilon _{\alpha \gamma}} - \frac{1}{2} \tensor{\Gamma}{^\epsilon _{\lambda \eta}}\tensor{R}{^\eta _{\gamma \alpha \xi}} y^\lambda y^\xi]\\
	&-&\frac{1}{4}\tensor{R}{^\epsilon _{\beta \alpha \xi}}y^\xi [(\tensor{R}{^\delta _{\gamma \lambda \epsilon}} +\tensor{R}{^\delta _{\epsilon \lambda \gamma}} + 2\frac{\partial \tensor{\Gamma}{^\delta _{\gamma \epsilon}}}{\partial x^\lambda})y^\lambda
	+ \tensor{\Gamma}{^\delta _{\nu \kappa}}(\tensor{R}{^\kappa _{\epsilon \mu \eta}} \tensor{\Gamma}{^\eta _{\lambda \gamma}}+\tensor{R}{^\kappa _{\gamma \mu \eta}} \tensor{\Gamma}{^\eta _{\lambda \epsilon}})y^\lambda y^\mu y^\nu]\\
	&-&[\tensor{\Gamma}{^\delta _{\epsilon \gamma}} - \frac{1}{2} \tensor{\Gamma}{^\delta _{\mu \kappa}}\tensor{R}{^\kappa _{\gamma \epsilon \nu}} y^\mu y^\nu][\tensor{\Gamma}{^\epsilon _{\alpha \beta}} - \frac{1}{2} \tensor{\Gamma}{^\epsilon _{\lambda \eta}}\tensor{R}{^\eta _{\beta \alpha \xi}} y^\lambda y^\xi]
\end{eqnarray*}

\begin{eqnarray*}
	\tensor{K}{^\delta_{(n+\alpha)\beta(n+\gamma)}}&=& -\frac{1}{2}\frac{\partial}{\partial x^{n+\gamma}}[\tensor{R}{^\delta _{\beta \alpha \lambda}} y^\lambda] -\frac{1}{4} \tensor{R}{^\delta _{\epsilon \gamma \mu}} \tensor{R}{^\epsilon _{\beta \alpha \nu}} y^\mu y^\nu 
\end{eqnarray*}

\begin{eqnarray*}
	\tensor{K}{^{(n+\delta)} _{(n+\alpha) \beta (n+\gamma)}}&=&
	-\frac{\partial}{\partial x^{(n+\gamma)}}
	[\tensor{\Gamma}{^\delta _{\alpha \beta}} -\frac{1}{2} \tensor{\Gamma}{^\delta _{\mu \lambda}}\tensor{R}{^\lambda _{\beta \alpha \nu}}y^\mu y^\nu] - \frac{1}{2} [\tensor{\Gamma}{^\delta _{\gamma \epsilon}} -\frac{1}{2} \tensor{\Gamma}{^\delta _{\mu \lambda}}\tensor{R}{^\lambda _{\epsilon \gamma \nu}}y^\mu y^\nu]\tensor{R}{^\epsilon _{\beta \alpha \kappa}}y^\kappa 
\end{eqnarray*}

\begingroup\makeatletter\def\f@size{9.7}\check@mathfonts
\def\maketag@@@#1{\hbox{\m@th\normalsize\normalfont#1}}%
\begin{eqnarray*}
	\tensor{K}{^\delta_{\alpha \beta \gamma}}&=& \frac{\partial}{\partial x^\beta}[\tensor{\Gamma}{^\delta _{\alpha \gamma}} +\frac{1}{2}(\tensor{R}{^\delta _{\gamma \kappa \mu}}\tensor{\Gamma}{^\kappa _{\nu \alpha}}+\tensor{R}{^\delta _{\alpha \kappa \mu}}\tensor{\Gamma}{^\kappa _{\nu \gamma}})y^\mu y^\nu]-\frac{\partial}{\partial x^\gamma}[\tensor{\Gamma}{^\delta _{\alpha \beta}} +\frac{1}{2}(\tensor{R}{^\delta _{\beta \kappa \mu}}\tensor{\Gamma}{^\kappa _{\nu \alpha}}+\tensor{R}{^\delta _{\alpha \kappa \mu}}\tensor{\Gamma}{^\kappa _{\nu \beta}})y^\mu y^\nu]\\
	&+&[\tensor{\Gamma}{^\delta _{\beta \epsilon}} +\frac{1}{2}(\tensor{R}{^\delta _{\epsilon \kappa \mu}}\tensor{\Gamma}{^\kappa _{\nu \beta}}+\tensor{R}{^\delta _{\beta \kappa \mu}}\tensor{\Gamma}{^\kappa _{\nu \epsilon}})y^\mu y^\nu][\tensor{\Gamma}{^\epsilon _{\alpha \gamma}} +\frac{1}{2}(\tensor{R}{^\delta _{\gamma \kappa \lambda}}\tensor{\Gamma}{^\kappa _{\xi \alpha}}+\tensor{R}{^\delta _{\alpha \kappa \lambda}}\tensor{\Gamma}{^\kappa _{\xi \gamma}})y^\lambda y^\xi]\\
	&-&[\tensor{\Gamma}{^\delta _{\gamma \epsilon}} +\frac{1}{2}(\tensor{R}{^\delta _{\epsilon \kappa \mu}}\tensor{\Gamma}{^\kappa _{\nu \gamma}}+\tensor{R}{^\delta _{\gamma \kappa \mu}}\tensor{\Gamma}{^\kappa _{\nu \epsilon}})y^\mu y^\nu][\tensor{\Gamma}{^\epsilon _{\alpha \beta}} +\frac{1}{2}(\tensor{R}{^\delta _{\beta \kappa \lambda}}\tensor{\Gamma}{^\kappa _{\xi \alpha}}+\tensor{R}{^\delta _{\alpha \kappa \lambda}}\tensor{\Gamma}{^\kappa _{\xi \beta}})y^\lambda y^\xi]\\
	&+&\frac{1}{4}\tensor{R}{^\delta _{\beta \epsilon \xi}}y^\xi[(\tensor{R}{^\epsilon _{\alpha \lambda \gamma}}+\tensor{R}{^\epsilon _{\gamma \lambda \alpha}}+2\frac{\partial \tensor{\Gamma}{^\epsilon _{\alpha \gamma}}}{\partial x^\lambda})y^\lambda +\tensor{\Gamma}{^\epsilon _{\nu \kappa}}(\tensor{R}{^\kappa _{\gamma \mu \eta}} \tensor{\Gamma}{^\eta _{\lambda \alpha}} + \tensor{R}{^\kappa _{\alpha \mu \eta}} \tensor{\Gamma}{^\eta _{\lambda \gamma}})y^\lambda y^\mu y^\nu]\\
	&-&\frac{1}{4}\tensor{R}{^\delta _{\gamma \epsilon \xi}}y^\xi[(\tensor{R}{^\epsilon _{\alpha \lambda \beta}}+\tensor{R}{^\epsilon _{\beta \lambda \alpha}}+2\frac{\partial \tensor{\beta}{^\epsilon _{\alpha \beta}}}{\partial x^\lambda})y^\lambda +\tensor{\beta}{^\epsilon _{\nu \kappa}}(\tensor{R}{^\kappa _{\beta \mu \eta}} \tensor{\beta}{^\eta _{\lambda \alpha}} + \tensor{R}{^\kappa _{\alpha \mu \eta}} \tensor{\beta}{^\eta _{\lambda \beta}})y^\lambda y^\mu y^\nu]
\end{eqnarray*}
\endgroup

\begingroup\makeatletter\def\f@size{8.9}\check@mathfonts
\def\maketag@@@#1{\hbox{\m@th\normalsize\normalfont#1}}%
\begin{eqnarray*}
	\tensor{K}{^{(n+\delta)}_{\alpha \beta \gamma}}&=& \frac{1}{2}\frac{\partial}{\partial x^\beta}[(\tensor{R}{^\delta _{\alpha \lambda \gamma}}+\tensor{R}{^\delta _{\gamma \lambda \alpha}}+2\frac{\partial \tensor{\Gamma}{^\delta _{\alpha \gamma}}}{\partial x^\lambda})y^\lambda + \tensor{\Gamma}{^\delta _{\nu \kappa}} (\tensor{R}{^\kappa _{\gamma \mu \eta}} \tensor{\Gamma}{^\eta _{\lambda \alpha}} + \tensor{R}{^\kappa _{\alpha \mu \eta}} \tensor{\Gamma}{^\eta _{\lambda \gamma}} )y^\lambda y^\mu y^\nu ]\\
	&-&\frac{1}{2}\frac{\partial}{\partial x^\gamma}[(\tensor{R}{^\delta _{\alpha \lambda \beta}}+\tensor{R}{^\delta _{\beta \lambda \alpha}}+2\frac{\partial \tensor{\Gamma}{^\delta _{\alpha \beta}}}{\partial x^\lambda})y^\lambda + \tensor{\Gamma}{^\delta _{\nu \kappa}} (\tensor{R}{^\kappa _{\beta \mu \eta}} \tensor{\Gamma}{^\eta _{\lambda \alpha}} + \tensor{R}{^\kappa _{\alpha \mu \eta}} \tensor{\Gamma}{^\eta _{\lambda \beta}} )y^\lambda y^\mu y^\nu ]\\
	&+& \frac{1}{2}[(\tensor{R}{^\delta _{\beta \lambda \epsilon}} + \tensor{R}{^\delta _{\epsilon \lambda \beta}} +2 \frac{\partial \tensor{\Gamma}{^\delta _{\beta \epsilon}}}{\partial x^\lambda})y^\lambda + \tensor{\Gamma}{^\delta _{\nu \kappa}} (\tensor{R}{^\kappa _{\epsilon \mu \eta}} \tensor{\Gamma}{^\eta _{\lambda \beta}} + \tensor{R}{^\kappa _{\beta \mu \eta}} \tensor{\Gamma}{^\eta _{\lambda \epsilon}} )y^\lambda y^\mu y^\nu]\dot{}\\
	&&\dot{}[\tensor{\Gamma}{^\epsilon _{\alpha \gamma}}+ \frac{1}{2}(\tensor{R}{^\epsilon _{\gamma \kappa \xi}} \tensor{\Gamma}{^\kappa _{\rho \alpha}} + \tensor{R}{^\epsilon _{\alpha \kappa \xi}} \tensor{\Gamma}{^\kappa _{\rho \gamma}} )y^\rho y^\xi]\\
	&-& \frac{1}{2}[(\tensor{R}{^\delta _{\gamma \lambda \epsilon}} + \tensor{R}{^\delta _{\epsilon \lambda \gamma}} +2 \frac{\partial \tensor{\Gamma}{^\delta _{\gamma \epsilon}}}{\partial x^\lambda})y^\lambda + \tensor{\Gamma}{^\delta _{\nu \kappa}} (\tensor{R}{^\kappa _{\epsilon \mu \eta}} \tensor{\Gamma}{^\eta _{\lambda \gamma}} + \tensor{R}{^\kappa _{\gamma \mu \eta}} \tensor{\Gamma}{^\eta _{\lambda \epsilon}} )y^\lambda y^\mu y^\nu]\dot{}\\
	&&\dot{}[\tensor{\Gamma}{^\epsilon _{\alpha \beta}}+ \frac{1}{2}(\tensor{R}{^\epsilon _{\beta \kappa \xi}} \tensor{\Gamma}{^\kappa _{\rho \alpha}} + \tensor{R}{^\epsilon _{\alpha \kappa \xi}} \tensor{\Gamma}{^\kappa _{\rho \beta}} )y^\rho y^\xi]\\
	&+& \frac{1}{2}[(\tensor{R}{^\epsilon _{\alpha \lambda \gamma}} + \tensor{R}{^\epsilon _{\gamma \lambda \alpha}} +2 \frac{\partial \tensor{\Gamma}{^\epsilon _{\alpha \gamma}}}{\partial x^\lambda})y^\lambda + \tensor{\Gamma}{^\epsilon _{\nu \kappa}} (\tensor{R}{^\kappa _{\gamma \mu \eta}} \tensor{\Gamma}{^\eta _{\lambda \alpha}} + \tensor{R}{^\kappa _{\alpha \mu \eta}} \tensor{\Gamma}{^\eta _{\lambda \gamma}} )y^\lambda y^\mu y^\nu][\tensor{\Gamma}{^\delta _{\epsilon \beta}}- \frac{1}{2}\tensor{R}{^\kappa _{\beta \epsilon \xi}} \tensor{\Gamma}{^\delta _{\rho \kappa}} y^\rho y^\xi]\\
	&-& \frac{1}{2}[(\tensor{R}{^\epsilon _{\alpha \lambda \beta}} + \tensor{R}{^\epsilon _{\beta \lambda \alpha}} +2 \frac{\partial \tensor{\Gamma}{^\epsilon _{\alpha \beta}}}{\partial x^\lambda})y^\lambda + \tensor{\Gamma}{^\epsilon _{\nu \kappa}} (\tensor{R}{^\kappa _{\beta \mu \eta}} \tensor{\Gamma}{^\eta _{\lambda \alpha}} + \tensor{R}{^\kappa _{\alpha \mu \eta}} \tensor{\Gamma}{^\eta _{\lambda \beta}} )y^\lambda y^\mu y^\nu][\tensor{\Gamma}{^\delta _{\epsilon \gamma}}- \frac{1}{2}\tensor{R}{^\kappa _{\gamma \epsilon \xi}} \tensor{\Gamma}{^\delta _{\rho \kappa}} y^\rho y^\xi]
\end{eqnarray*}
\endgroup

\begin{eqnarray*}
	\tensor{K}{^\delta _{\alpha \beta (n+\gamma)}} &=& \frac{1}{2}\frac{\partial}{\partial x^\beta}[\tensor{R}{^\delta _{\alpha \gamma \lambda}} y^\lambda] - \frac{\partial}{\partial x^{(n+\gamma)}}[\tensor{\Gamma}{^\delta _{\alpha \beta}}- \frac{1}{2}(\tensor{R}{^\delta _{\beta \kappa \mu}} \tensor{\Gamma}{^\kappa _{\lambda \alpha}} + \tensor{R}{^\delta _{\alpha \kappa \mu}} \tensor{\Gamma}{^\kappa _{\lambda \beta}} )y^\lambda y^\mu]\\
	&+&\frac{1}{2}\tensor{R}{^\delta _{\beta \epsilon \nu}}y^\nu[\tensor{\Gamma}{^\epsilon _{\gamma \alpha}}- \frac{1}{2}\tensor{R}{^\kappa _{\alpha \gamma \lambda}} \tensor{\Gamma}{^\epsilon _{\mu \kappa}} y^\lambda y^\mu]\\
	&+&\frac{1}{2}\tensor{R}{^\epsilon _{\alpha \gamma \nu}}y^\nu[\tensor{\Gamma}{^\delta _{\beta \epsilon}}+ \frac{1}{2}(\tensor{R}{^\delta _{\epsilon \kappa \mu}} \tensor{\Gamma}{^\kappa _{\lambda \beta}} + \tensor{R}{^\delta _{\beta \kappa \mu}} \tensor{\Gamma}{^\kappa _{\lambda \epsilon}} )y^\lambda y^\mu]\\
	&-&\frac{1}{2}\tensor{R}{^\delta _{\epsilon \gamma \nu}}y^\nu[\tensor{\Gamma}{^\epsilon _{\alpha \beta}}+ \frac{1}{2}(\tensor{R}{^\epsilon _{\beta \kappa \mu}} \tensor{\Gamma}{^\kappa _{\lambda \alpha}} + \tensor{R}{^\epsilon _{\alpha \kappa \mu}} \tensor{\Gamma}{^\kappa _{\lambda \beta}} )y^\lambda y^\mu]
\end{eqnarray*}

\begin{eqnarray*}
	\tensor{K}{^{(n+\delta)} _{\alpha \beta (n+\gamma)}} &=& \frac{\partial}{\partial x^{\beta}}[\tensor{\Gamma}{^\delta _{\alpha \gamma}}- \frac{1}{2}\tensor{R}{^\kappa _{\alpha \gamma \lambda}} \tensor{\Gamma}{^\delta _{\mu \kappa}}y^\lambda y^\mu]\\
	&-&\frac{1}{2}\frac{\partial}{\partial x^{(n+\gamma)}}[(\tensor{R}{^\delta _{\alpha \lambda \beta}}+\tensor{R}{^\delta _{\beta \lambda \alpha}}+2\frac{\partial \tensor{\Gamma}{^\delta _{\alpha \beta}}}{\partial x^\lambda})y^\lambda + \tensor{\Gamma}{^\delta _{\nu \kappa}} (\tensor{R}{^\kappa _{\beta \mu \eta}} \tensor{\Gamma}{^\eta _{\lambda \alpha}} + \tensor{R}{^\kappa _{\alpha \mu \eta}} \tensor{\Gamma}{^\eta _{\lambda \beta}} )y^\lambda y^\mu y^\nu ]\\
	&+& \frac{1}{2}\tensor{R}{^\epsilon _{\alpha \gamma \rho}}y^\rho[(\tensor{R}{^\delta _{\beta \lambda \epsilon}} + \tensor{R}{^\delta _{\epsilon \lambda \beta}} +2 \frac{\partial \tensor{\Gamma}{^\delta _{\beta \epsilon}}}{\partial x^\lambda})y^\lambda + \tensor{\Gamma}{^\delta _{\nu \kappa}} (\tensor{R}{^\kappa _{\epsilon \mu \eta}} \tensor{\Gamma}{^\eta _{\lambda \beta}} + \tensor{R}{^\kappa _{\beta \mu \eta}} \tensor{\Gamma}{^\eta _{\lambda \epsilon}} )y^\lambda y^\mu y^\nu]\\
	&+& [\tensor{\Gamma}{^\epsilon _{\gamma \alpha}}- \frac{1}{2}\tensor{R}{^\kappa _{\alpha \gamma \lambda}} \tensor{\Gamma}{^\epsilon _{\mu \kappa}} y^\lambda y^\mu][\tensor{\Gamma}{^\delta _{\epsilon \beta}}- \frac{1}{2}\tensor{R}{^\kappa _{\beta \epsilon \xi}} \tensor{\Gamma}{^\delta _{\rho \kappa}} y^\rho y^\xi]\\
	&-& [\tensor{\Gamma}{^\epsilon _{\alpha \beta}}+ \frac{1}{2} (\tensor{R}{^\epsilon _{\beta \kappa \nu}} \tensor{\Gamma}{^\kappa _{\rho \alpha}} + \tensor{R}{^\epsilon _{\alpha \kappa \nu}} \tensor{\Gamma}{^\kappa _{\rho \beta}} )y^\rho y^\nu][\tensor{\Gamma}{^\delta _{\epsilon \gamma}}- \frac{1}{2}\tensor{R}{^\kappa _{\epsilon \gamma \xi}} \tensor{\Gamma}{^\delta _{\mu \kappa}} y^\mu y^\xi]
\end{eqnarray*}

\begin{eqnarray*}
	\tensor{K}{^\delta _{\alpha (n+\beta) \gamma}} &=&
	\frac{\partial}{\partial x^{(n+\beta)}}[\tensor{\Gamma}{^\delta _{\alpha \gamma}}+ \frac{1}{2}(\tensor{R}{^\delta _{\gamma \kappa \mu}} \tensor{\Gamma}{^\kappa _{\lambda \alpha}} + \tensor{R}{^\delta _{\alpha \kappa \mu}} \tensor{\Gamma}{^\kappa _{\lambda \gamma}} )y^\lambda y^\mu]-
	\frac{1}{2}\frac{\partial}{\partial x^\gamma} [\tensor{R}{^\delta _{\alpha \beta \lambda}} y^\lambda]\\
	&+&\frac{1}{2}\tensor{R}{^\delta _{ \epsilon \beta \nu}}y^\nu[\tensor{\Gamma}{^\epsilon _{\gamma \alpha}}+ \frac{1}{2}(\tensor{R}{^\epsilon _{\gamma \kappa \mu}} \tensor{\Gamma}{^\kappa _{\lambda \alpha}}+\tensor{R}{^\epsilon _{\alpha \kappa \mu}} \tensor{\Gamma}{^\kappa _{\lambda \gamma}}) y^\lambda y^\mu]\\
	&-&\frac{1}{2}\tensor{R}{^\epsilon _{\alpha \beta \nu}}y^\nu[\tensor{\Gamma}{^\delta _{\gamma \epsilon}}+ \frac{1}{2}(\tensor{R}{^\delta _{\epsilon \kappa \mu}} \tensor{\Gamma}{^\kappa _{\lambda \gamma}}+\tensor{R}{^\delta _{\gamma \kappa \mu}} \tensor{\Gamma}{^\kappa _{\lambda \epsilon}}) y^\lambda y^\mu]
	\\
	&-&\frac{1}{2}\tensor{R}{^\delta _{\gamma \epsilon \nu}}y^\nu[\tensor{\Gamma}{^\epsilon _{\beta \alpha}}-\frac{1}{2}\tensor{R}{^\kappa _{\alpha \beta \lambda}} \tensor{\Gamma}{^\epsilon _{\mu \kappa}} y^\lambda y^\mu]
\end{eqnarray*}

\begin{eqnarray*}
	\tensor{K}{^{(n+\delta)} _{\alpha (n+\beta) \gamma }} &=&
	\frac{1}{2}\frac{\partial}{\partial x^{(n+\beta)}}[(\tensor{R}{^\delta _{\alpha \lambda \gamma}}+\tensor{R}{^\delta _{\gamma \lambda \alpha}}+2\frac{\partial \tensor{\Gamma}{^\delta _{\alpha \gamma}}}{\partial x^\lambda})y^\lambda + \tensor{\Gamma}{^\delta _{\nu \kappa}} (\tensor{R}{^\kappa _{\gamma \mu \eta}} \tensor{\Gamma}{^\eta _{\lambda \alpha}} + \tensor{R}{^\kappa _{\alpha \mu \eta}} \tensor{\Gamma}{^\eta _{\lambda \gamma}} )y^\lambda y^\mu y^\nu ]\\
	&-&\frac{\partial}{\partial x^{\gamma}}[\tensor{\Gamma}{^\delta _{\alpha \beta}}- \frac{1}{2}\tensor{R}{^\kappa _{\alpha \beta \lambda}} \tensor{\Gamma}{^\delta _{\mu \kappa}}y^\lambda y^\mu]\\
	&-& \frac{1}{2}\tensor{R}{^\epsilon _{\alpha \beta \rho}}y^\rho[(\tensor{R}{^\delta _{\gamma \lambda \epsilon}} + \tensor{R}{^\delta _{\epsilon \lambda \gamma}} +2 \frac{\partial \tensor{\Gamma}{^\delta _{\gamma \epsilon}}}{\partial x^\lambda})y^\lambda + \tensor{\Gamma}{^\delta _{\nu \kappa}} (\tensor{R}{^\kappa _{\epsilon \mu \eta}} \tensor{\Gamma}{^\eta _{\lambda \gamma}} + \tensor{R}{^\kappa _{\gamma \mu \eta}} \tensor{\Gamma}{^\eta _{\lambda \epsilon}} )y^\lambda y^\mu y^\nu]\\
	&-& [\tensor{\Gamma}{^\epsilon _{\beta \alpha}}- \frac{1}{2}\tensor{R}{^\kappa _{\alpha \beta \lambda}} \tensor{\Gamma}{^\epsilon _{\mu \kappa}} y^\lambda y^\mu][\tensor{\Gamma}{^\delta _{\epsilon \gamma}}- \frac{1}{2}\tensor{R}{^\kappa _{\gamma \epsilon \xi}} \tensor{\Gamma}{^\delta _{\rho \kappa}} y^\rho y^\xi]\\
	&+& [\tensor{\Gamma}{^\epsilon _{\alpha \gamma}}+ \frac{1}{2} (\tensor{R}{^\epsilon _{\gamma \kappa \nu}} \tensor{\Gamma}{^\kappa _{\rho \alpha}} + \tensor{R}{^\epsilon _{\alpha \kappa \nu}} \tensor{\Gamma}{^\kappa _{\rho \gamma}} )y^\rho y^\nu][\tensor{\Gamma}{^\delta _{\epsilon \beta}}- \frac{1}{2}\tensor{R}{^\kappa _{\epsilon \beta \xi}} \tensor{\Gamma}{^\delta _{\mu \kappa}} y^\mu y^\xi]
\end{eqnarray*}

\begingroup\makeatletter\def\f@size{9.7}\check@mathfonts
\def\maketag@@@#1{\hbox{\m@th\normalsize\normalfont#1}}%
\begin{eqnarray*}
	\tensor{K}{^\delta _{\alpha (n+\beta) (n+\gamma)}} &=&
	\frac{1}{2}\frac{\partial}{\partial x^{(n+\beta)}}[\tensor{R}{^\delta _{\gamma \alpha \lambda}}y^\lambda]-\frac{1}{2}\frac{\partial}{\partial x^{(n+\gamma)}} [\tensor{R}{^\delta _{\beta \alpha \lambda}} y^\lambda]+\frac{1}{4}\tensor{R}{^\delta _{ \epsilon \beta \nu}}\tensor{R}{^\epsilon _{ \alpha \gamma \mu}}y^\mu y^\nu
	-\frac{1}{4}\tensor{R}{^\delta _{ \epsilon \gamma \nu}}\tensor{R}{^\epsilon _{ \alpha \beta \mu}}y^\mu y^\nu
\end{eqnarray*}
\endgroup

\begin{eqnarray*}
	\tensor{K}{^{(n+\delta)} _{\alpha (n+\beta) \gamma }} &=&\frac{\partial}{\partial x^{(n+\beta)}}[\tensor{\Gamma}{^\delta _{\alpha \gamma}}- \frac{1}{2}\tensor{R}{^\kappa _{\alpha \gamma \lambda}} \tensor{\Gamma}{^\delta _{\mu \kappa}}y^\lambda y^\mu]-\frac{\partial}{\partial x^{(n+\gamma)}}[\tensor{\Gamma}{^\delta _{\alpha \beta}}- \frac{1}{2}\tensor{R}{^\kappa _{\alpha \beta \lambda}} \tensor{\Gamma}{^\delta _{\mu \kappa}}y^\lambda y^\mu]\\
	&+&[\tensor{\Gamma}{^\epsilon _{\alpha \gamma}}+ \frac{1}{2} (\tensor{R}{^\epsilon _{\gamma \kappa \nu}} \tensor{\Gamma}{^\kappa _{\rho \alpha}} + \tensor{R}{^\epsilon _{\alpha \kappa \nu}} \tensor{\Gamma}{^\kappa _{\rho \gamma}} )y^\rho y^\nu][\tensor{\Gamma}{^\delta _{\epsilon \beta}}- \frac{1}{2}\tensor{R}{^\kappa _{\epsilon \beta \xi}} \tensor{\Gamma}{^\delta _{\mu \kappa}} y^\mu y^\xi]\\
	&-&[\tensor{\Gamma}{^\epsilon _{\alpha \beta}}+ \frac{1}{2} (\tensor{R}{^\epsilon _{\beta \kappa \nu}} \tensor{\Gamma}{^\kappa _{\rho \alpha}} + \tensor{R}{^\epsilon _{\alpha \kappa \nu}} \tensor{\Gamma}{^\kappa _{\rho \beta}} )y^\rho y^\nu][\tensor{\Gamma}{^\delta _{\epsilon \gamma}}- \frac{1}{2}\tensor{R}{^\kappa _{\epsilon \gamma \xi}} \tensor{\Gamma}{^\delta _{\mu \kappa}} y^\mu y^\xi]
\end{eqnarray*}
where $\tensor{\Gamma}{^\alpha _{\beta \gamma}}$, $\tensor{R}{^\delta _{\alpha \beta \gamma}}$ are the Christoffel symbols and Riemann curvature tensor of the $n$-dimensional base manifold $(M,g)$, respectively, and, inkeeping with our previous convention for Section~\ref{section3}, the~greek indices $\{\alpha, \beta,...\}= \{1,2, \cdots n\}$ and the latin indices $\{a, b,...\}= \{1,2, \cdots 2n\}$.

It should be noted that if the base manifold is flat, i.e.,~in an appropriate coordinate system $\tensor{R}{^\delta _{\alpha \beta \gamma}}=0$ and $\tensor{\Gamma}{^\alpha _{\beta \gamma}}=0$ then $\bar{\Gamma}\tensor{}{^i_{jk}}=0$ and $\tensor{K}{^i _{j h k}} = 0$. Therefore, a~flat base manifold induces a flat Sasaki tangent bundle. In~this case the ``geodesics'' of the flat tangent space are all trivially produced by lifting the ``geodesics'' of the base~manifold.

% The following MDPI journals use author-date citation: Arts, Econometrics, Economies, Genealogy, Humanities, IJFS, JRFM, Laws, Religions, Risks, Social Sciences. For those journals, please follow the formatting guidelines on http://www.mdpi.com/authors/references
% To cite two works by the same author: \citeauthor{ref-journal-1a} (\citeyear{ref-journal-1a}, \citeyear{ref-journal-1b}). This produces: Whittaker (1967, 1975)
% To cite two works by the same author with specific pages: \citeauthor{ref-journal-3a} (\citeyear{ref-journal-3a}, p. 328; \citeyear{ref-journal-3b}, p.475). This produces: Wong (1999, p. 328; 2000, p. 475)

%=====================================
% References, variant B: external bibliography
%=====================================
%\externalbibliography{yes}
%\bibliography{your_external_BibTeX_file}

\begin{thebibliography}{999}
	% Reference 13
	\bibitem[Farnes (2018)]{Farnes}
	Farnes, J.S. A unifying theory of dark energy and dark matter: Negative masses and matter creation within a modified $\Lambda$CDM framework. \textit{Astron. Astrophys.} \textbf{2018}, \textit{620}, A92.
	% Reference 9
	\bibitem[Nadler, E. O. et al. (2020)]{Fermilab}
	Nadler, E.O.; Wechsler, R.H.; Bechtol, K.; Mao, Y.Y.; Green, G.; Drlica-Wagner, A.; McNanna, M.; Mau, S.; Pace, A.B.;  Simon, J.D.;~et~al. Milky Way Satellite Census. II. Galaxy--Halo Connection Constraints Including the Impact of the Large Magellanic Cloud. \textit{Astrophys. J.} \textbf{2020}, \textit{893}, 48.
	% Reference 10
	\bibitem[Xiangxiang (2018)]{mediator}
	Ren, X.; Zhao, L.; Abdukerim, A.; Chen, X.; Chen, Y.; Cui, X.; Fang, D.; Fu, C.; Giboni, K.;  Giuliani, F.;~et~al. Constraining Dark Matter Models with a Light Mediator at the PandaX-II Experiment. \textit{Phys. Rev. Lett.} \textbf{2018}, \textit{121}, 021304. 
	% Reference 11
	\bibitem[Dror (2020)]{data}
	Dror, J.A.; Elor, G.; McGehee, R. Directly Detecting Signals from Absorption of Fermionic Dark Matter. \textit{\mbox{Phys. Rev. Lett.}} \textbf{2020}, \textit{124}, 181301.
	% Reference 12
	\bibitem[Hsueh (2020)]{Royal}
	Hsueh, J.W.; Enzi, W.; Vegetti, S.; Auger, M.W.; Fassnacht, C.D.; Despali, G.; Koopmans, L.V.; McKean, J.P. SHARP---VII. New constraints on the dark matter free-streaming properties and substructure abundance from gravitationally lensed quasars. \textit{Mon. Not. R. Astron. Soc.} \textbf{2020}, \textit{492}, 3047--3059.
	
	% Reference 32
	\bibitem[Riess (1998)]{Riess1}
	Riess, A.G.; Filippenko, A.V.; Challis, P.; Clocchiatti, A.; Diercks, A.; Garnavich, P.M.; Gilliland; R.L.; Hogan,~C.J.; Jha, S.;  Kirshner, R.P.;~et~al. Observational evidence from supernovae for an
	accelerating universe and a cosmological constant. %The duplicate has been removed.
	\textit{Astron. J.} \textbf{1998}, \textit{116}, 1009–1038.
	
	% Reference 33
	\bibitem[Rigault (2018)]{arxiv}
	Rigault, M.; Brinnel, V.; Aldering, G.; Antilogus, P.; Aragon, C.; Bailey, S.; Baltay, C.; Barbary, K.; Bongard,~S.;  Boone, K.; et~al. Strong Dependence of Type Ia
	Supernova Standardization on the Local Specific Star
	Formation Rate. {\it arXiv} \textbf{2018},  arXiv:1806.03849.
	
	% Reference 34
	\bibitem[Martinelli (2019)]{Symmetry}
	Martinelli, M.; Tutusaus, I. CMB tensions with
	low-redshift H0 and S8 measurements: Impact of a
	redshift-dependent type-Ia supernovae intrinsic
	luminosity. \textit{Symmetry} \textbf{2019}, \textit{11}, 986.
	
	% Reference 35
	\bibitem[Valentino (2020)]{Valentino1}
	Di Valentino, E.; Gariazzo, S.; Mena, O.; Vagnozzi, S. Soundness of Dark Energy Properties. \textit{J. Cosmol. Astropart. Phys.} \textbf{2020},  doi:10.1088/1475-7516/2020/07/045.
	
	% Reference 36
	\bibitem[Dhawan (2018)]{Leib}
	Dhawan, S.; Jha S.W.;  Leibundgut, B. Measuring
	the Hubble constant with Type Ia supernovae as
	near-infrared standard candles. \textit{Astron. Astrophys.} \textbf{2018}, \textit{609}, A72.
	
	% Reference 37
	\bibitem[Leib (2005)]{Leib2}
	Leibundgut, B.; Blondin, S. Evidence for dark energy from Type Ia supernovae.
	\textit{Nucl. Phys. B Proc. Suppl.} \textbf{2005}, \textit{138}, 10--15.
	
	% Reference 38
	\bibitem[Krauss (2007)]{Krauss}
	Krauss, L. M.;  Jones-Smith, K.;  Huterer, D. Dark energy, a cosmological constant, and type Ia supernovae. \textit{New J. Phys.} \textbf{2007},  \textit{9}, 141 %We added all the authors and the title.
	
	% Reference 39
	\bibitem[Perlmutter (2003)]{Saul1}
	Perlmutter, S. Supernovae, Dark Energy, and the Accelerating Universe. \textit{Phys. Today} \textbf{2003}, \textit{56}, 53. 
	
	% Reference 40
	\bibitem[Riess (2004)]{Riess2}
	Riess, A.G.  Type Ia supernova discoveries at z > 1 from the Hubble Space Telescope: Evidence for past deceleration and constraints on dark energy evolution. \textit{Astrophys. J.} \textbf{2004}, \textit{607}, 665--687.
	
	% Reference 41
	\bibitem[Abbott (2019)]{Abbott}
	Abbott, T.M.C.; Allam, S.; Andersen, P.; Angus, C.; Asorey, J.; Avelino, A.; Avila, S.; Bassett, B.A.; Bechtol, K.;  Bernstein, G.M.;~et~al. First Cosmology Results using Type Ia Supernovae from the Dark Energy Survey: Constraints on Cosmological Parameters. \textit{ Astrophys. J. Lett.} \textbf{2019}, \textit{872}, L30.
	
	% Reference 42
	\bibitem[Kang (2019)]{Yijung}
	Kang, Y.; Lee, Y.W.; Kim, Y.L.; Chung, C.; Ree, C.H. Early-type Host Galaxies of Type Ia Supernovae. II. Evidence for Luminosity Evolution in Supernova Cosmology. {\it arXiv} \textbf{2019}, arXiv:1912.04903.
	
	% Reference 43
	\bibitem[Nielsen (2016)]{Sarkar}
	Nielsen, J.T.; Guffanti, A.; Sarkar, S. Marginal evidence for cosmic acceleration from Type Ia supernovae. \textit{Sci.~Rep.} \textbf{2016}, \textit{6}, 35596.
	
	% Reference 44
	\bibitem[Valentino (2017)]{Valentino2}
	Di Valentino, E.; Melchiorri, A.; Linder, E.V.; Silk, J. Constraining dark energy dynamics in extended parameter space. \textit{Phys. Rev. D.} \textbf{2017}, \textit{96}, 023523.
	
	% Reference 45
	\bibitem[Leib (2018)]{Leib3}
	Leibundgut. B.; Sullivan, M. Type Ia supernova cosmology. \textit{Space Sci. Rev.} \textbf{2018}, \textit{214}, 57.
	
	% Reference 46
	\bibitem[Perlmutter (2018)]{Saul2}
	Perlmutter, S.; Aldering, G.; Goldhaber, G.; Knop, R.A.; Nugent, P.; Castro, P.G.; Deustua, S.; Fabbro, S.; Goobar, A.;  Groom, D.E.;~et~al. Measurements of $\Omega$ and $\Lambda$ from 42 high-redshift supernovae. \textit{Astrophys. J.} \textbf{1999}, \textit{517}, 565.
	
	% Reference 48
	\bibitem[Velten (2018)]{Velten}
	Velten, H.; Gomes, S.; Busti, V.C. Gauging the cosmic acceleration with recent type Ia supernovae data sets. \textit{Phys. Rev. D.} \textbf{2018}, \textit{97}, 083516.
	
	% Reference 49
	\bibitem[Hoscheit (2018)]{Barger}
	Hoscheit, B.L.; Barger, A.J. The KBC Void: Consistency with Supernovae Type Ia and the Kinematic SZ Effect in a $\Lambda$LTB Model. \textit{Astrophys. J.} \textbf{2018}, \textit{854}, 46.
	
	% Reference 50
	\bibitem[Gariazzo (2017)]{Diamanti}
	Gariazzo, S.; Escudero, M.; Diamanti, R.; Mena, O. Cosmological searches for a noncold dark matter component. \textit{Phys. Rev. D} \textbf{2017}, \textit{96}, 043501.
	
	% Reference 51
	\bibitem[Valentino (2020)]{Valentino3}
	Di Valentino, E.; Melchiorri, A.; Mena, O.;
	Vagnozzi, S. Nonminimal dark sector physics and
	cosmological tensions. \textit{Phys. Rev. D} \textbf{2020}, \textit{101}, 063502.
	
	% Reference 52
	\bibitem[Collet (2018)]{GR}
	Collett, T.E.; Oldham, L.J.; Smith, R.J.; Auger, M.W.; Westfall, K.B.; Bacon, D.; Nichol, R.C.; Masters, K.L.; Koyama, K.; van den Bosch, R. A Precise Extragalactic Test of General Relativity. \textit{Science} \textbf{2018}, \textit{360}, 1342--1346.
	
	% Reference 1
	\bibitem[Hartle, J. B.(2003)]{Hartle}
	Hartle, J.B. {\em Gravity: An Introduction to Einstein's General Relativity}; Addison-Wesley: Boston, MA, USA,  2003
	
	% Reference 26
	\bibitem[Borowiec (2007)]{Godlowski}
	Borowiec, A.; Godłowski, W.; Szydłowski, M. Dark Matter and Dark Energy as Effects of Modified Gravity. \textit{Int. J. Geom. Methods Mod. Phys.} \textbf{2007}, \textit{4}, 183--196
	% Reference 27
	\bibitem[Salucci (1995)]{Salucci}
	Persic, M.; Salucci, P.; Stel, F. The Universal Rotation Curve of Spiral Galaxies: I. the Dark Matter Connection. \textit{Mon. Not. R. Astron. Soc.} \textbf{1995}, \textit{283}, 27--47.
	% Reference 28
	\bibitem[Harko (2012)]{Harko1}
	Harko, T.; Lobo, F. Geodesic deviation, Raychaudhuri equation, and tidal forces in modified gravity with an arbitrary curvature-matter coupling. \textit{Phys. Rev. D.} \textbf{2012}, \textit{86}, 124034.
	% Reference 29
	\bibitem[Harko (2014)]{Harko2}
	Harko, T.; Lobo, F. Generalized Curvature-Matter Couplings in Modified Gravity. \textit{Galaxies} \textbf{2014}, \textit{2}, 410--465.
	
	% Reference 2
	\bibitem[Sasaki (1958)]{Sas-58}
	Sasaki, S. On the differential geometry of tangent bundles of Riemannian manifolds. \textit{Tohoku Math. J.} \textbf{1958}, \textit{10},~338--354.
	% Reference 14
	\bibitem[Mukhanov (2017)]{Mukhanov}
	Chamseddine, A.H.; Mukhanov, V. Resolving Cosmological Singularities. \textit{J. Cosmol. Astropart. Phys.} \textbf{2017}, {\it 2017}, 009.
	% Reference 15
	\bibitem[Ball (2005)]{Nature}
	Ball, P. Dark matter highlights extra dimensions. \textit{Nature} \textbf{2005}, doi:10.1038/news050829-18.
	% Reference 16
	\bibitem[Harko (2011)]{Question}
	Kahil, M.; Harko, T. Is dark matter an extra-dimensional effect? \textit{Mod. Phys. Lett. A} \textbf{2011}, \textit{24}, 667--682.
	% Reference 17
	\bibitem[Coimbra (2013)]{Brazil}
	Coimbra Araújo, C.H.; da Rocha, R. Gravity with Extra Dimensions and Dark Matter Interpretation: A~Straightforward Approach. \textit{ISRN High Energy Phys.} \textbf{2013}, {\it 2013}, 713508.
	% Reference 18
	\bibitem[Tanabashi (2018)]{Group}
	Tanabashi, M. Particle Data Group. \textit{Phys. Rev. D} \textbf{2018}, \textit{98}, 030001. 
	
	% Reference 58
	\bibitem[Miller (1981)]{Miller}
	Miller, K.S. On the Inverse of the Sum of Matrices. \textit{Math. Mag.} \textbf{1981}, \textit{54}, 67--72.
	
	% Reference 31
	\bibitem[Capozziello (2006)]{Capoz}
	Capozziello, S.; Cardone, V. F.;  Troisi, A. Dark Energy and Dark Matter as Curvature Effects? \textit{J. Cosmol. Astropart. Phys.} \textbf{2006}, \textit{2006}, doi:10.1088/1475-7516/2006/08/001.
	% Reference 3
	\bibitem[Raychaudhuri (1955)]{Ray}
	Raychaudhuri, A. Relativistic Cosmology. \textit{Phys. Rev.} \textbf{1955}, \textit{98}, 1123. 
	% Reference 4
	\bibitem[Hawk. Ell. (1973)]{Hawk}
	Hawking, S.; Ellis, G. \textit{The Large Scale Structure of Space-Time}; Cambridge Monographs on Mathematical Physics; Cambridge University Press: Cambridge, UK, 1973.
	% Reference 5
	\bibitem[Sengupta (2007)]{Sengupta}
	Kar, S.; Sengupta, S. The Raychaudhuri equations: A brief review. \textit{Pramana J. Phys.} \textbf{2007}, \textit{69}, 49--76.
	% Reference 6
	\bibitem[Alexiou (2018)]{Alexiou}
	Stavrinos, P. C.; Alexiou, M. Raychaudhuri equation in the Finsler--Randers space-time and generalized scalar-tensor theories. \textit{Int. J. Geom. Methods Mod. Phys.} \textbf{2018}, \textit{15}, 1850039.
	
	% Reference 19
	\bibitem[Moreshi (2015)]{Moreshi}
	Moreshi,O.; Boero, E.; Gallo E.; Gesser, F. Dark matter description by non-conventional energy-momentum tensor. In {\it AIP Conference Proceedings}; American Institute of Physics: College Park, MD, USA, 2015; Volume~1647, pp. 35--43.
	% Reference 20
	\bibitem[Peebles (2000)]{Peebles}
	Peebles, P.J.E. Fluid Dark Matter. \textit{ Astrophys. J.} \textbf{2000}, \textit{534}, L127.
	% Reference 21
	\bibitem[Arbey (2006)]{Alexandre}
	Arbey, A. Dark Fluid: A complex scalar field to unify dark energy and dark matter. \textit{Phys. Rev. D} \textbf{2006}, \textit{74},~043516.
	
	% Reference 22
	\bibitem[Kuzmichev (2012)]{Kuzmichev}
	Kuzmichev, V.; Kuzmichev, V. Two-component perfect fluid in FRW universe. \textit{Acta Phys. Pol. Ser. B} \textbf{2012}, \textit{43},~1899--1910.
	% Reference 23
	\bibitem[Alvarenga (2016)]{Rodolfo}
	Alvarenga, F.; Fracalossi, R.; Freitas, R. C.; Gonçalves, S. Classical and quantum cosmology with two perfect fluids: Stiff matter and radiation. \textit{Gen. Relativ. Gravit.} \textbf{2016}, \textit{49}, 136.
	% Reference 24
	\bibitem[Farrando (1990)]{Ferrando}
	Ferrando, J.J.; Morales, J.A.; Portilla, M. Two-perfect fluid interpretation of an energy tensor. \textit{Gen. Relativ. Gravit.} \textbf{1990}, \textit{22}, 1021--1032
	% Reference 25
	\bibitem[Oliveira (1989)]{Oliveira}
	Oliveira, S.R. Model of two perfect fluids for an anisotropic and homogeneous universe. \textit{Phys. Rev. D} \textbf{1989}, \textit{40}, 3976
	
	% Reference 59
	\bibitem[Sotiriou (2010)]{Sotiriou}
	Sotiriou, T.P.; Faraoni, V.  {\it F (R)} Theor. Gravity. \textit{Rev. Mod. Phys.} \textbf{2010}, \textit{82}, 451--497
	
	% Reference 60
	\bibitem[De Felice (2010)]{Antonio}
	De Felice, A.; Tsujikawa, S. {\it F (R)} Theories. \textit{Living Rev. Relativ.} \textbf{2010}, \textit{13}, 3.
	
	% Reference 53
	\bibitem[Capozziello (2009)]{Capoz2}
	Capozziello, S.; De Laurentis, M.; Francaviglia, M.; Mercadante, S. From Dark Energy \& Dark Matter to Dark Metric. \textit{Found. Phys.} \textbf{2009}, \textit{39}, 1161--1176.
	
	% Reference 54
	\bibitem[Nojiri (2008)]{Nojiri}
	Nojiri, S.; Odintsov, S.D. Dark energy, inflation and dark matter from modified F(R) gravity. {\it arXiv} \textbf{2008}, arXiv:0807.0685
	
	% Reference 55
	\bibitem[Tupper (1990)]{Tupper}
	Tupper, B.O.J. Conformally Ricci‐flat viscous fluids.
	\textit{J. Math. Phys.} \textbf{1990}, \textit{31}, 1704.
	
	% Reference 56
	\bibitem[Hansraj (2013)]{Hindawi}
	Hansraj, S.; Govinder, K.; Mewalal, N. Conformal Mappings in Relativistic Astrophysics. \textit{J. Appl. Math.} \textbf{2013}, \textit{2013}, 196385.
	
	% Reference 57
	\bibitem[De Felice (1990)]{Felice}
	de Felice, F.; Clarke, C.J.S. {\em Relativity on Curved Manifolds}; Cambridge Monographs on Mathematical Physics; Cambridge University Press: Cambridge, UK, 1990.
	
	% Reference 61
	\bibitem[Carroll (2004)]{Carroll}
	Carroll, S.M. {\em Spacetime and Geometry: An Introduction to General Relativity}; Addison-Wesley: Boston, MA, USA, 2004.
	
	% Reference 62
	\bibitem[Liddle (2000)]{Liddle}
	Liddle, A.; Lyth, D. {\em Cosmological Inflation and Large-Scale Structure}; Cambridge University Press: Cambridge, UK, 2000.
	
	% Reference 30
	\bibitem[Rund (2014)]{Rund}
	Rund, H. {\em The Differential Geometry of Finsler Spaces}; Springer:  Berlin/Heidelberg, Germany, %we confirm this new piece of information
	1959.
	
	% Reference 7
	\bibitem[Bejan (2018)]{French}
	Bejan, C.; Gül, I. Sasaki metric on the tangent bundle of a Weyl manifold. \textit{Publ. Inst. Math.} \textbf{2018}, \textit{103}, 25--32.
	% Reference 8
	\bibitem[Kowalski (1971)]{German}
	Kowalski, O. Curvature of the Induced Riemannian Metric on the Tangent Bundle of a Riemannian Manifold. \textit{J. Reine Angew. Math.} \textbf{1971}, \textit{250}, 124--129.
	
\end{thebibliography}

%%%%%%%%%%%%%%%%%%%%%%%%%%%%%%%%%%%%%%%%%%
%% optional
%\sampleavailability{Samples of the compounds ...... are available from the authors.}

%% for journal Sci
%\reviewreports{\\
%Reviewer 1 comments and authors’ response\\
%Reviewer 2 comments and authors’ response\\
%Reviewer 3 comments and authors’ response
%}

%%%%%%%%%%%%%%%%%%%%%%%%%%%%%%%%%%%%%%%%%%
\end{document}